\documentclass[aps,prd,reprint,superscriptaddress,nofootinbib,preprintnumbers
]{revtex4-1}

\usepackage[utf8]{inputenc}
\usepackage{amsmath,amssymb}
\usepackage[separate-uncertainty=true]{siunitx}
\usepackage{graphicx}
\usepackage[usenames,dvipsnames]{xcolor}
\usepackage{subcaption}
\usepackage[colorlinks=true
,urlcolor=blue
,anchorcolor=blue
,citecolor=blue
,filecolor=blue
,linkcolor=blue
,menucolor=blue
,linktocpage=true
,pdfproducer=medialab
,pdfa=true
]{hyperref}
\usepackage{booktabs}
\usepackage{tabularx}
\usepackage{xspace}
\usepackage[normalem]{ulem}
\usepackage{slashed}
\usepackage{bbold}
\usepackage{wasysym}
\usepackage{mathrsfs} 
\usepackage{multirow}
\usepackage[compat=1.0.0]{tikz-feynman}
\usepackage{orcidlink}
\usepackage{placeins}
\usepackage{caption}
\makeatletter
\long\def\@makecaption#1#2{%
  \vskip\abovecaptionskip
  \begingroup
    \small\upshape
    \leftskip=0pt\relax\rightskip=0pt\relax\parfillskip=0pt plus 1fil\relax
    \noindent\@make@capt@title{#1}{#2}\par
  \endgroup
  \vskip\belowcaptionskip
}
\makeatother
\usepackage{tablefootnote}
\usepackage{threeparttable}

\usepackage{appendix}
\usepackage{bm}


\allowdisplaybreaks

\graphicspath{{graphics/}}

\newcommand{\ie}{i.e.,~}
\newcommand{\eg}{e.g.,~}

\newcommand{\pb}{{\ensuremath\rm pb}\xspace}
\newcommand{\fb}{{\ensuremath\rm fb}\xspace}

\DeclareSIUnit{\pb}{pb}
\DeclareSIUnit{\fb}{fb}
\AtBeginDocument{
\heavyrulewidth=.08em
\lightrulewidth=.05em
\cmidrulewidth=.03em
\belowrulesep=.65ex
\belowbottomsep=0pt
\aboverulesep=.4ex
\abovetopsep=0pt
\cmidrulesep=\doublerulesep
\cmidrulekern=.5em
\defaultaddspace=.5em

\newcolumntype{C}{>{\centering\arraybackslash}X}
\newcolumntype{b}{C}
\newcolumntype{s}{>{\hsize=.6\hsize}C}
\newcolumntype{R}{>{\raggedleft\arraybackslash}X}
}

\newcommand{\tauh}{\ensuremath{\tau_\textrm{h}}\xspace}
\newcommand{\taul}{\ensuremath{\tau_\ell}\xspace}
\newcommand{\tauhtauh}{\ensuremath{\tau_\textrm{h}\tau_\textrm{h}}\xspace}
\newcommand{\taultauh}{\ensuremath{\tau_\ell\tau_\textrm{h}}\xspace}
\newcommand{\phicp}{\ensuremath{\phi_{CP}}\xspace}
\newcommand{\CP}{\ensuremath{C\!P}\xspace}

\makeatletter
\AtBeginDocument{\if@filesw\immediate\write\@auxout{\string\citation{apsrev41Control}}\fi}
\makeatother

\begin{document}
\date{\today}
\rightline{}
\title{\vspace{6mm}{\Large TauPolaris: reconstructing tau lepton polarimetric vectors with conditional normalizing flows}\vspace{3mm}}

\author{Daniel Winterbottom\orcidlink{0000-0003-4582-150X}}
\email{d.winterbottom15@imperial.ac.uk}
\affiliation{Imperial College London,
Department of Physics, Blackett Laboratory, SW7 2BW, United Kingdom}

\author{Lucas Russell\orcidlink{0000-0002-6502-2185}}
\email{lucas.russell@cern.ch}
\affiliation{Imperial College London,
Department of Physics, Blackett Laboratory, SW7 2BW, United Kingdom}

\renewcommand{\abstractname}{\texorpdfstring{\vspace{0.5cm}}{} Abstract}

{
\begin{abstract}
    \vspace{0.5cm}
The kinematics of tau lepton decay products depend on the tau spin, giving access to the spin
correlations and \CP structure of the process that produced them. The optimal spin observable is
the polarimetric vector, which points along the most likely direction of the tau spin.
Determining it requires the momenta of the neutrinos produced in the tau decays, which escape
detection and must therefore be inferred.
We present \texttt{TauPolaris}, a tool for reconstructing tau polarimetric vectors by estimating
the undetected neutrino momenta with a conditional normalizing flow. Rather than performing a
point regression, the method models the full conditional density of the neutrino kinematics,
from which spin observables and their uncertainties are estimated.
The resolution of the reconstructed spin observables improves on that obtained from networks
trained with a mean-squared-error loss. Using simulated LHC proton-proton collisions, including
detector resolution effects, we demonstrate the method in three applications. For tau leptons
produced in Higgs boson decays, we show that the presence of quantum entanglement can be
distinguished from its absence with a significance of at least $4.3\sigma$ at the High-Luminosity LHC. We demonstrate a
36\% improvement in the sensitivity to \CP violation in $H\rightarrow\tau\tau$ decays.
Finally, we introduce new variables for suppressing the $Z\rightarrow\tau\tau$ background in Higgs
boson searches.

\end{abstract}

\maketitle
}

\section{Introduction}
\label{sec:introduction}
Tau leptons have short lifetimes compared with the scale of typical collider detectors,
and they tend to decay within a few mm of the primary collision point.
This means we never observe the $\tau$ itself, only the particles it decays into.
Tau leptons decay either leptonically, into an electron or muon and two neutrinos,
or hadronically into a neutrino and a collection of charged and neutral hadrons (mainly pions).

While these decays make $\tau$ leptons much more challenging to reconstruct,
they also open up new possibilities for measurements that are not possible for the lighter leptons.
Namely, the angular correlations between the decay products are sensitive to the spin orientation,
so we can use the decay topology to indirectly constrain the $\tau$ lepton spin.
In principle, if we could reconstruct all of the $\tau$ decay products precisely,
we could measure an optimal variable for determining the $\tau$ spin,
usually referred to as the $\tau$'s polarimetric vector.

Measuring $\tau$ spin has several uses:
searches for \CP violation in both $H\rightarrow\tau\tau$~\cite{Kramer:1993jn,CMS:2021sdq,ATLAS:2022akr,CMS:2026hvv}
and $Z/\gamma^*\rightarrow\tau\tau$~\cite{Fabbrichesi:2024wcd,LoChiatto:2024dmx} processes,
which may help to explain the observed baryon asymmetry of the universe;
tests of quantum entanglement at colliders~\cite{Fabbrichesi:2022ovb,Altakach:2022ywa,Fabbrichesi:2024wcd,Zhang:2025mmm};
measurements of higher-dimensional effective operators,
including those which modify the $\tau$ lepton's electric and magnetic dipole
moments~\cite{Fabbrichesi:2024wcd,LoChiatto:2024dmx};
precise measurements of the $\tau$ polarization in $Z$ boson decays,
which determine the effective weak mixing angle from the $\tau$ couplings alone
and thereby test the lepton universality of the weak neutral current~\cite{CMS:2023mgq};
and as a handle for suppressing background processes.

However, reconstructing the polarimetric vectors is very challenging,
since the neutrinos produced in $\tau$ decays are not detected.
Current analysis methods instead rely on approximate constructions for defining spin-sensitive
observables, which limits the experimental precision and is not easily transferable between
different processes and energy regimes.

Several methods have previously been proposed for reconstructing neutrinos in $\tau\tau$ events,
combining kinematic constraints and, in some cases, information on the $\tau$ lifetime, with
assumptions to fix the remaining
unknowns~\cite{ELLIS1988221,Elagin:2010aw,Bianchini:2016yrt,Kalinowski:2025vci,Konar:2016wbh,Gripaios:2012th,Cherepanov:2018npf,Bhardwaj:2016lcu}.
However, none of these methods use all the available constraints, and several rely on approximations,
such as assuming the neutrinos are collinear with the visible decay products, that bias the
angular correlations on which $\tau$ spin measurements critically depend.
A complete treatment would also require the construction of a likelihood function including the
detector response for all measured inputs, which cannot be parametrized exactly in analytic form
and can be slow to evaluate.
Machine learning (ML) instead learns these effects directly from simulation, without the need to
parametrize them, and is fast to evaluate once trained.

In this work we present a new ML method for reconstructing $\tau$ lepton polarimetric vectors based on
conditional normalizing flows.
The flow learns the full conditional density of the neutrino kinematics rather than a single point
estimate, which brings several advantages.
Sampling from the density reproduces the generator-level distributions,
while the most probable configuration provides an accurate per-event estimate.
The spread of the sampled density also gives a per-event uncertainty estimate.
Crucially, because the flow models the joint density of the neutrino momentum components,
any configuration drawn from it respects the correlations between them.
A regression algorithm trained on a per-component mean-squared-error objective instead returns the conditional
mean of each component, which need not correspond to any allowed configuration:
where the event kinematics permit two distinct solutions, the mean lies between them,
biasing the neutrino toward the visible $\tau$ direction
and degrading the angular correlations that carry the spin information.

Normalizing flows have previously been used for neutrino reconstruction in
Refs.~\cite{Leigh:2022lpn,Raine:2023fko}, for neutrinos produced in semileptonic top-quark decays.
The topologies of these decays are very different from those of $\tau$ decays:
the neutrinos originate from typically on-shell $W$ boson decays,
and thus tend to be well separated from the other decay products.
In contrast, the $\tau$ lepton has a small mass in comparison to its typical energy,
so its decay products are collimated into a narrow cone in the detector.
This makes an exact measurement of the angular correlations between the visible decay products and
the neutrino very challenging, and these are exactly what spin measurements depend on.
Ref.~\cite{Zhang:2025mmm} employed an alternative approach using a diffusion model to learn the conditional density for neutrinos from a subset of $Z\rightarrow\tau\tau$ decays. 
However, this setup does not provide a tractable density, precluding a direct estimation of the most probable neutrino configuration; the neutrino momenta are instead obtained by stochastic sampling, which is generally slower for diffusion models than for normalizing flows~\cite{Krause:2024avx}.

In this work we focus on the normalizing flow method, introducing several developments tailored to $\tau$ spin measurements:
(i) including additional variables sensitive to the finite $\tau$ lifetime
(\ie the displacements of the visible decay products);
(ii) an orthonormal coordinate system for parametrizing the neutrino momenta,
defined with respect to the visible $\tau$ momentum so as to align closely with the basis in which
the spin correlations are expressed; (iii) per-event estimates of the spin observables and their uncertainties obtained by averaging over samples drawn from the learned density, together with a gradient-based maximization routine for determining the most probable neutrino configuration;
and (iv) a transformer-based embedding network for processing the conditioning variables.\footnote{Ref.~\cite{Zhang:2025mmm} also employed (iv).}
Furthermore, we use the estimated neutrinos to compute the $\tau$ polarimetric vectors, which, as noted above, are the optimal spin-analyzing observables.

We demonstrate the performance of the resulting tool, \texttt{TauPolaris}\footnote{The code is publicly available at \url{https://github.com/danielwinterbottom/TauPolaris}.}, in three benchmark
studies: entanglement measurements at the LHC, improved measurements of \CP violation in
$H\rightarrow\tau\tau$ decays, and new variables for separating $H\rightarrow\tau\tau$ and
$Z\rightarrow\tau\tau$ events.

This paper is organized as follows.
Section~\ref{sec:theory} introduces the formalism used to describe $\tau$ spin correlations at
colliders, and Sec.~\ref{sec:simulations} describes the simulated samples, the modeling of
detector effects, and the reconstruction of the $\tau$ decay products.
The \texttt{TauPolaris} models are presented in Sec.~\ref{sec:models},
together with a comparison of their performance with that of a stand-alone transformer regressor.
Sections~\ref{sec:entanglement},~\ref{sec:cp} and~\ref{sec:spin} present the three benchmark
applications described above, and we summarize in Sec.~\ref{sec:summary}.

\section{Tau Spin Measurements at Colliders} 
\label{sec:theory}
The spin direction of a $\tau$ lepton cannot be determined exactly on an 
event-by-event basis. However, the polarimetric vector, $\boldsymbol{h}$, constructed 
from the momenta of the $\tau$ lepton's decay products, gives the most likely 
direction of the $\tau$ spin and provides maximal sensitivity to it. 

Correlations between the polarimetric vectors of a $\tau$ lepton pair encode the spin
correlations and \CP structure of the process that produced them.
We define the spin state of ditau pairs in a right-handed orthonormal basis $\{{\bf n},{\bf r},{\bf k}\}$, constructed in the ditau rest frame, where $\bf k$ points in the direction of the $\tau^+$, and:
\begin{equation}\label{eq:coord}
    {\bf n}=\frac{1}{\sin{\Theta}}({\bf p}\times{\bf k}), \qquad
    {\bf r}=\frac{1}{\sin{\Theta}}({\bf p}-{\bf k}\cos{\Theta}),
\end{equation}
with ${\bf p} = (0,0,-1)$ and ${\bf p}\cdot{\bf k}=\cos{\Theta}$.
The density matrix describing the spin state of the ditau system can be written as
\begin{widetext}
\begin{equation}
\rho = \frac{1}{4}\left[\mathbb{1}\otimes\mathbb{1} + \sum_{i}B_i^+(\sigma_i\otimes\mathbb{1}) + \sum_{j}B_j^-(\mathbb{1}\otimes\sigma_j) + \sum_{i,j}C_{ij}(\sigma_i\otimes\sigma_j)\right],
\end{equation}
\end{widetext}
where $i,j\in\{{\bf n},{\bf r},{\bf k}\}$, and $\sigma_i$ are the Pauli matrices. The $B_i^{\pm}$ coefficients encode the polarization of the corresponding $\tau^\pm$ lepton, while the $C_{ij}$ matrix describes the spin correlations.

The presence of \CP violation is indicated if $C_{ij}\neq C_{ji}$ for any $i \neq j$. For a Higgs boson decay, the \CP nature modifies the $\mathbf{C}$ matrix depending on the value of a mixing angle, $\alpha$:
\begin{equation}
\label{eqn:C_higgs}
\mathbf{C} = 
\begin{pmatrix}
\cos{2\alpha} & \sin{2\alpha} & 0 \\
-\sin{2\alpha} & \cos{2\alpha} & 0 \\
0 & 0 & -1
\end{pmatrix},
\end{equation}
and the $\tau$ leptons are unpolarized ($B_i^{\pm}=0~\forall~i$)~\cite{Altakach:2022ywa}.
The mixing angle depends on the \CP-even and \CP-odd $\tau$ lepton Yukawa couplings, $y_{\tau}$ and $\tilde{y}_{\tau}$~\cite{CMS:2026hvv}:
\begin{equation}
\alpha=\arctan\left({\frac{\tilde{y}_{\tau}}{y_{\tau}}}\right).
\end{equation}

Quantum entanglement is quantified by the ``concurrence''~\cite{Wootters:1997id}, $\mathcal{C}$, which is defined as 
\begin{equation}
\label{eqn:concurrence}
\mathcal{C} = \max(0,\lambda_1-\lambda_2-\lambda_3-\lambda_4),    
\end{equation}
where $\lambda_i=\sqrt{r_i}$ and $r_i$ are the eigenvalues in descending order of the matrix $\rho(\sigma_2\otimes\sigma_2)\rho^*(\sigma_2\otimes\sigma_2)$. The concurrence can take values of $0\leq \mathcal{C}\leq 1$, where $\mathcal{C}=1$ indicates a maximally entangled state and $\mathcal{C}=0$ a separable (non-entangled) system. 
We also define a variable $m_{12}$ which can serve as a Bell non-locality witness~\cite{Fabbrichesi:2024wcd}. This is defined by the eigenvalues $m_i$ of the matrix $\mathbf{M}=\mathbf{C}\mathbf{C}^{T}$, which are ordered in decreasing order $m_1\geq m_2\geq m_3$:
\begin{equation}
\label{eqn:m12}
    m_{12} = m_1 + m_2.
\end{equation}
A value of $m_{12}>1$ signals a violation of Bell's inequality.
The $\tau$ leptons from Higgs boson decays are expected to have $\mathcal{C}=1$ and $m_{12}=2$ independent of $\alpha$.

\section{Simulations and Event Reconstruction}
\label{sec:simulations}
We generate samples of ditau events produced in LHC proton-proton collisions at a center-of-mass energy of $13.6\,\text{TeV}$ using \texttt{Pythia} (version 8.312)~\cite{Bierlich:2022pfr} for the hard process, parton showering, and fragmentation, as well as the decay of the $\tau$ leptons. The underlying event is modeled using the CP5 tune~\cite{CMS:2019csb}.
Hadronic $\tau$ decays (\tauh) to $\pi^\mp\nu$, $\pi^\mp\pi^0\nu$, $\pi^\mp\pi^0\pi^0\nu$, $\pi^\mp\pi^\mp\pi^\pm\nu$, and $\pi^\mp\pi^\mp\pi^\pm\pi^0\nu$, as well as leptonic decays (\taul) to $e^\mp\nu\bar\nu$ and $\mu^\mp\nu\bar\nu$, are considered.
The short-hand used for hadronic decays is $N_{\pi^\pm}\pi^\pm N_{\pi^0}\pi^0$ where $N_{\pi^\pm}$ ($N_{\pi^0}$) are the number of charged (neutral) pions in the decay. 
 
Samples are produced for both gluon-fusion-induced Higgs boson events ($pp\rightarrow H\rightarrow\tau\tau$) and Drell--Yan events ($pp\rightarrow Z/\gamma^*\rightarrow\tau\tau$). 
The Higgs boson sample is generated with the $\tau$ lepton spin correlations disabled in \texttt{Pythia}, and a reweighting method is then applied to model arbitrary components of the spin density matrix, $\rho$. This is used to model different \CP properties for the $H\rightarrow\tau\tau$ decay: \CP-even ($\alpha=0^\circ$), \CP-odd ($\alpha=90^\circ$), and mixed-\CP ($\alpha=45^\circ$). 

The reweighting follows the formalism implemented in \texttt{TauSpinner}~\cite{Przedzinski:2018ett}, modified to allow arbitrary values of the elements of $\mathbf{B}^{\pm}$ and $\mathbf{C}$ to be specified.\footnote{The polarimetric-vector calculation, based on the \texttt{TAUOLA} formalism~\cite{Jadach:1990mz,Jezabek:1991qp,Jadach:1993hs,Davidson:2010rw}, and the reweighting described here were reimplemented in Python for this work.} Events are reweighted according to
\begin{widetext}
\begin{equation}
    W_T = 1 + \sum_{i}B_i^{+}\cos\theta_i^{+} + \sum_{j}B_j^{-}\cos\theta_j^{-} + \sum_{i,j}C_{ij}\cos\theta_i^{+}\cos\theta_j^{-},
\end{equation}
\end{widetext}
where $\cos\theta_i^{\pm}$ is the projection of the polarimetric vector $\boldsymbol{h}^{\pm}$ of the $\tau^{\pm}$, normalized to unit length, onto the axis $i\in\{\mathbf{n},\mathbf{r},\mathbf{k}\}$, and $B_i^{\pm}$ and $C_{ij}$ are as defined in Sec.~\ref{sec:theory}.

The detector simulation is based on the \texttt{Delphes} fast simulation framework (version 3.5)~\cite{deFavereau:2013fsa} using the CMS detector configuration,  with some additional detector-level smearing performed by hand, as described below.

The \texttt{Delphes} setup does not account for the smearing of the directions and impact parameters of tracks left in the detector by charged pions, muons, and electrons. We include this smearing as follows. Firstly, we smear the primary interaction point by sampling a Gaussian for each of the $x$, $y$, and $z$ components independently. The resolutions of the $x$/$y$ and $z$ components are taken as $5\,\mu\text{m}$ and $29\,\mu\text{m}$, respectively, using the values provided in the supplementary material of Ref.~\cite{CMS:2021sdq}. We additionally smear the $\phi$, $\cot\theta$, transverse impact parameter $d_0$, and longitudinal impact parameter $d_z$ of each track, using resolutions parametrized as a function of transverse momentum $p_{\text{T}}$ and pseudorapidity $\eta$ provided in Ref.~\cite{CMS:2014pgm} (Figure~15).

For $\tau$ lepton decays to $\pi^\mp\pi^\mp\pi^\pm\nu$ ($3\pi^\pm0\pi^0$) and $\pi^\mp\pi^\mp\pi^\pm\pi^0\nu$ ($3\pi^\pm1\pi^0$), the secondary decay vertex is reconstructed from the three charged pion tracks. Each track is represented as a straight line, with a reference point on the track constructed from the smeared $d_0$ and $d_z$ values as $(d_0\sin\phi, -d_0\cos\phi, d_z)$, and direction given by the smeared $\phi$ and $\cot\theta$. The secondary vertex is then estimated as the global least-squares solution minimizing the sum of squared distances from the three track lines.\footnote{Strictly, charged particles follow helical trajectories in the magnetic field of the detector. However, for the track lengths and momenta considered here, the curvature is sufficiently small that the tracks can be well approximated as straight lines.} The reconstructed secondary vertex position is expressed relative to the smeared primary vertex, and the vector from the primary to the secondary vertex gives an estimate of the $\tau$ flight direction.

The leptonically decaying $\tau$ leptons are reconstructed from muon and electron candidates identified by \texttt{Delphes}.
For hadronically decaying $\tau$ leptons, the visible decay products are reconstructed following an emulation of the CMS hadron-plus-strips (HPS) algorithm~\cite{CMS:2018jrd}. Strips representing $\pi^0$ candidates are formed by iteratively clustering photons within $p_\text{T}$-dependent windows in $\eta$ and $\phi$, and retaining those with $p_\text{T} > 2.5\,\text{GeV}$. In the $\tau^\mp \to \pi^\mp \pi^0 \pi^0 \nu$  ($1\pi^\pm2\pi^0$) decay mode, the photons from the two $\pi^0$s typically overlap in the detector and are clustered into a single strip object, as is the case in CMS data. The $\tau$ lepton candidates are then formed from combinations of one or three charged hadrons with zero or one strip, subject to mass window and signal cone requirements following Ref.~\cite{CMS:2018jrd}; the highest-$p_\text{T}$ candidate passing all requirements is selected. 
Both the leptonic and hadronic $\tau$ candidates are required to have $p_\text{T} > 20\,\text{GeV}$.

While the HPS algorithm does not distinguish between the $1\pi^\pm1\pi^0$ ($\tau^\mp \to \pi^\mp\pi^0\nu$) and $1\pi^\pm2\pi^0$ decay modes, CMS employs an additional multivariate classifier to separate these two cases~\cite{CMS-DP-2020-041}. Since this classifier is not straightforward to replicate in our simulation, we instead randomly assign candidates reconstructed with a single strip into the $1\pi^\pm1\pi^0$ or $1\pi^\pm2\pi^0$ class by sampling according to the decay-mode migration probabilities reported in Ref.~\cite{CMS-DP-2020-041}, conditioned on the true generator-level decay mode. In both cases the four-momentum of the reconstructed strip and the assigned decay mode are used as input variables for the algorithms described in Sec.~\ref{sec:models}.

\section{Model Architectures and Training}
\label{sec:models}

\subsection{Inputs and Datasets}

The neutrino kinematics are estimated using a separate right-handed orthonormal coordinate system for each $\tau$ lepton candidate, constructed from the visible $\tau$ lepton momentum. The basis $\{\bm{n_i}, \bm{r_i}, \bm{k_i}\}$ is defined with $\bm{k_i}$ pointing in the visible $\tau_i$ direction and $\bm{r_i}$ and $\bm{n_i}$ defined by Eq.~(\ref{eq:coord}).
This choice was found to yield better resolution on the spin observables than Cartesian coordinates, as the basis is closely related to the orthonormal set based on the full $\tau$ momentum described in Sec.~\ref{sec:theory}.
Separate models are trained for the fully hadronic (\tauhtauh) and semileptonic (\taultauh) ditau decay channels, since \taul decays produce an additional neutrino relative to \tauh decays.
In the case of \tauh decays, the neutrino three-momentum components are estimated. In the case of \taul decays, the summed three-momentum components of the electron/muon neutrino $\nu_\ell$ and $\tau$ lepton neutrino $\nu_\tau$ are targeted, as well as the invariant mass of the two neutrino system $m\left(\bm{\nu_{\tau}}+\bm{\nu_{\ell}}\right)$.
The regressed quantities are summarized in Table~\ref{tab:targets}. The $\tau$ leptons in the ditau pair are referred to as candidate 1 and candidate 2, with the corresponding estimated neutrino three-momentum vector referred to as $\bm{\nu_1}$ or $\bm{\nu_2}$ respectively.
There are therefore 6 outputs for the \tauhtauh models, and 7 outputs for the \taultauh models. These are collectively denoted by the vector $\bm{\nu}\in \mathbb{R}^{D}$ (where $D$ is the number of outputs).

\begin{table*}[htbp]
  \centering
  \setlength{\tabcolsep}{12pt}
  \caption{Neutrino kinematics estimated in the \tauhtauh and \taultauh channels. The neutrino three-momenta are estimated in orthonormal coordinates  $\{\bm{n_i}, \bm{r_i}, \bm{k_i}\}$ defined from the visible $\tau_i$ direction. In the \taultauh channel, the invariant mass of the $\nu_\ell$ and $\nu_\tau$ is also regressed.}
  \label{tab:targets}
  \begin{tabular}{l c c}
    \toprule
    Target & \tauhtauh & \taultauh \\
    \midrule
    $\bm{\nu_1}$ (coordinates $\{\bm{n_1}, \bm{r_1}, \bm{k_1}\})$ & $\bm{\nu_{\tau_1}}$ & $\bm{\nu_{\tau_1}} + \bm{\nu_{\ell1}} $ \\
    $\bm{\nu_2}$ (coordinates $\{\bm{n_2}, \bm{r_2}, \bm{k_2}\})$ & $\bm{\nu_{\tau_2}}$ & $\bm{\nu_{\tau_2}}$ \\
    $\nu_1$ mass  & -- & $m\left(\bm{\nu_{\tau_1}}+\bm{\nu_{\ell1}}\right)$ \\
    \bottomrule
  \end{tabular}
\end{table*}

Reconstructed event properties are used by the models presented in this work to constrain the kinematics of the undetected neutrinos.
These input features include the four-momenta of all visible $\tau$ decay products. In the $\tauh$ case, these comprise up to three charged pions (labeled $\pi^\pm_1$, $\pi^\pm_2$, $\pi^\mp_3$) and up to one reconstructed $\pi^0$ strip. In the \taul decay case, the charged lepton $\ell^\pm$ is the only visible decay product. 
The impact parameter and secondary vertex position vectors, as well as the missing transverse momentum are also included.
Finally, flags are used to determine the type of decay, by specifying the number of $\pi^\pm$ and $\pi^0$s in \tauh decays and the light lepton flavor in \taul decays. 
The full set of input features is summarized in Table~\ref{tab:input_features}; there are a total of 52 input variables for the \tauhtauh model and 35 inputs for the \taultauh model.

\begin{table*}[htbp]
  \centering
  \setlength{\tabcolsep}{7pt}
  \caption{Input features. The variables are separated into different groups based on the object represented. The number of instances of each set of features in the \taultauh and \tauhtauh channels is indicated in the  columns on the right. There are a total of 52 inputs in the \tauhtauh and 35 in the \taultauh channel.}
  \label{tab:input_features}
  \begin{tabular}{l l c c c}
  \toprule
    Object & Variables & $\textrm{N}_\textrm{features}$ & $\tauhtauh$ & $\taultauh$ \\
    \midrule
    \tauh visible decay products ($\pi^\pm_1, \pi^\pm_2, \pi^\mp_3, \pi^0$) & $(p_x, p_y, p_z, E)$ & 16 & $\times2$ & $\times1$ \\
    \tauh impact parameter & $d_x, d_y, d_z$ & 3 & $\times2$ & $\times1$ \\
    \tauh secondary decay vertex & $x, y, z$ & 3 & $\times2$ & $\times1$ \\
    \tauh decay mode flags & has-$\pi^0$, N$_\pi^0$, is-3-prong & 3 & $\times2$ & $\times1$ \\
    \taul visible decay products ($\ell^\pm$) & $(p_x, p_y, p_z, E)$ & 4 & --- & $\times1$ \\
    \taul impact parameter & $d_x, d_y, d_z$ & 3 & --- & $\times1$ \\
    \taul lepton flavor flag & is-$\mu$-or-$e$ & 1 & --- & $\times1$ \\
    Missing transverse momentum & $p_x, p_y$ & 2 & $\times1$ & $\times1$ \\
    \midrule
    Total features & & & 52 & 35 \\
    \bottomrule
  \end{tabular}
\end{table*}

The inputs are standardized via z-score normalization, with the mean and standard deviation computed from the training dataset, and subsequently used to rescale the training, validation and evaluation datasets. 
The simulated gluon-fusion-induced Higgs boson samples introduced above are used for training, with the sum of the \CP-even and \CP-odd \texttt{TauSpinner} weights applied. In the \tauhtauh channel, 51 million events are used for training, and 6 million for validation. In the \taultauh channel, 17 million events are used for training and 2 million are used for validation. The datasets used for testing are statistically independent.

\subsection{Conditional Flow Architecture}

\begin{figure*}[!htbp]
\includegraphics[width=\linewidth]{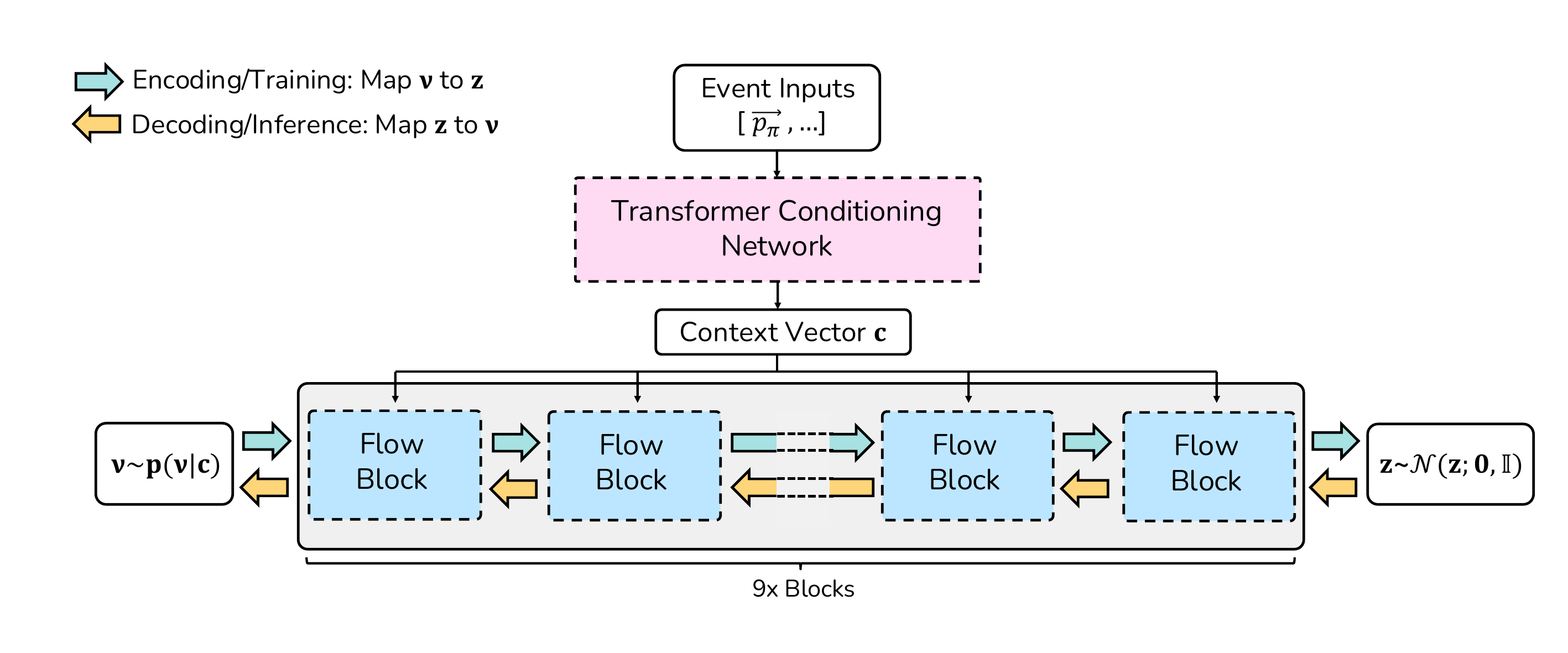}
\caption{Overview of the conditional normalizing flow architecture. The model maps the probability density of the neutrino kinematics $p(\bm{\nu} \mid  \mathbf{c})$, conditional on a context vector $\mathbf{c}$, to a $D$--dimensional Gaussian distribution $\mathcal{N}(\mathbf{z};\mathbf{0},\mathbb{I})$. A total of nine invertible conditional flow blocks are used for the transformation. The context vector is a summary of the event learned by a transformer network.  }
\label{fig:flow}
\end{figure*}

The conditional normalizing flow models the probability density of the neutrino kinematics $p(\bm{\nu}   \mid  \mathbf{c})$, conditional on a context vector $\mathbf{c}$, which is a learned summary of the event.
The flow expresses this non-trivial density using a series of invertible transformations $f$, mapping $\bm{\nu}$ to a latent variable $\mathbf{z} = f(\bm{\nu}   \mid \mathbf{c})$, which follows a $D$--dimensional Gaussian distribution $\mathcal{N}(\mathbf{z};\mathbf{0},\mathbb{I})$. The estimated neutrino density is then determined by the change of variables formula:
\begin{equation} \label{eq:changevars}
      \log p(\bm{\nu}  \mid \mathbf{c}) = \log \mathcal{N}\!\bigl(\mathbf{z} ;\, \mathbf{0}, \mathbb{I}\bigr) + \log \left| \det J_f(\bm{\nu}  \mid  \mathbf{c}) \right|
\end{equation}
where $J_f$ is the Jacobian of the transformation $f$.
An overview of the conditional normalizing flow architecture employed is shown in Fig.~\ref{fig:flow}.
The model is implemented using the \texttt{nflows} package~\cite{nflows} in \texttt{PyTorch}~\cite{torch}.
The hyperparameters described below were chosen using the Bayesian optimization package \texttt{Optuna}~\cite{optuna}.

\begin{figure*}[!htbp]
\includegraphics[width=0.95\linewidth]{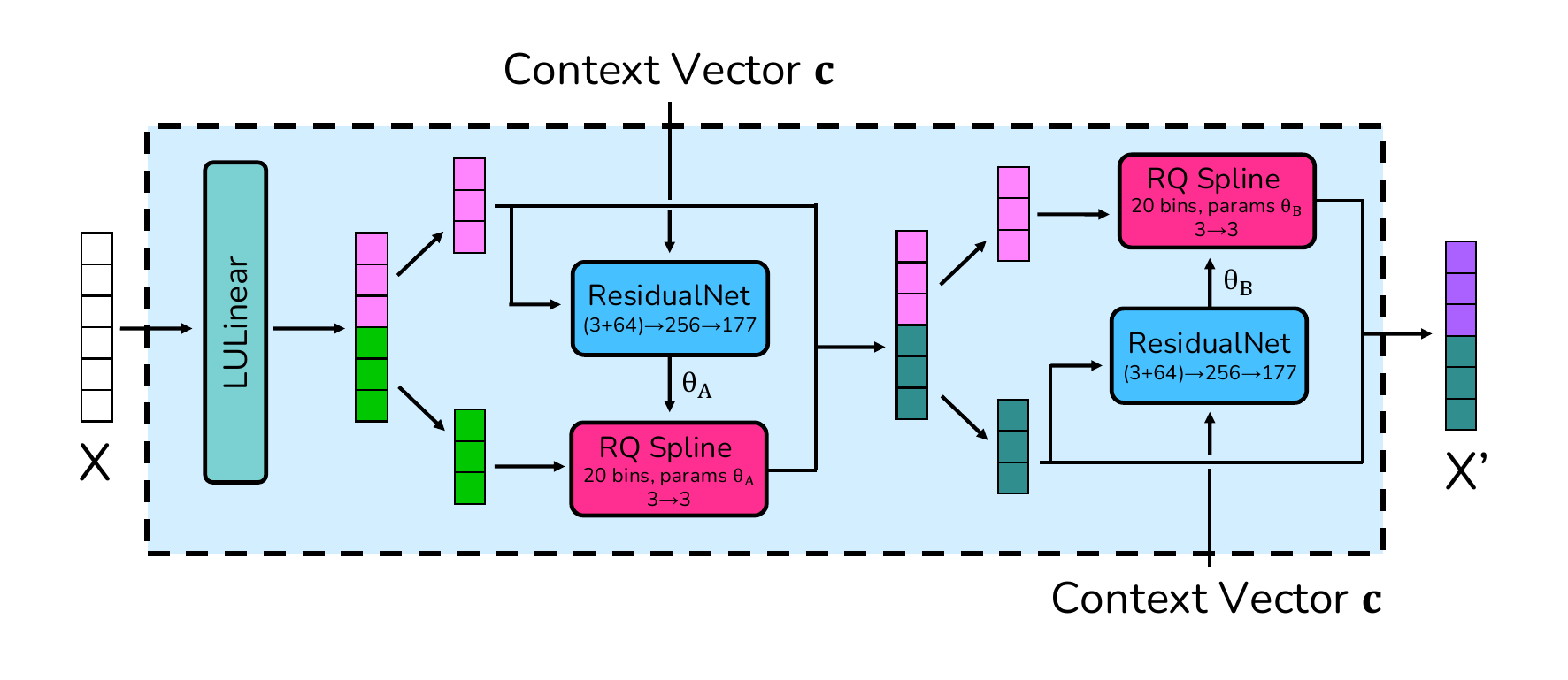}
\caption{Diagram of a conditional flow block, which is a fully invertible operation. Each block (except for the first) begins with an LU-decomposed linear mixing layer. Two coupling transformations are then performed on alternating halves of the input dimensions. In each coupling transformation, a monotonic piecewise rational-quadratic spline is applied to half of the input dimensions. The spline parameters ($\theta_A/\theta_B$) are determined by a residual network trained using the other half of the input dimensions and the context vector $\mathbf{c}$.}
\label{fig:flowblock}
\end{figure*}

The flow transformation consists of nine conditional flow blocks, depicted in Fig.~\ref{fig:flowblock}.
Each block begins with an LU-decomposed linear mixing layer~\cite{LUlinear} (except for the first block), followed by two successive rational quadratic spline coupling transformations  \cite{splineflows}, acting on alternating halves of the dimensions.
In each coupling transformation, a monotonic piecewise rational-quadratic spline (with 20 bins in the $[-4,4]$ range) is applied element-wise to half of the dimensions, and the other half are passed through unchanged. The parameters $\theta_A/\theta_B$ of the spline are determined with a residual network~\cite{resnet} with GELU activation~\cite{GELU} taking the unchanged half of the dimensions and the context vector as inputs. Each coupling transformation has an analytically computable inverse, allowing the mapping from $\mathbf{z}\to\bm{\nu}$ during inference, as well as $\bm{\nu}\to\mathbf{z}$ during training.

\begin{figure}[!htbp]
\includegraphics[width=\linewidth]{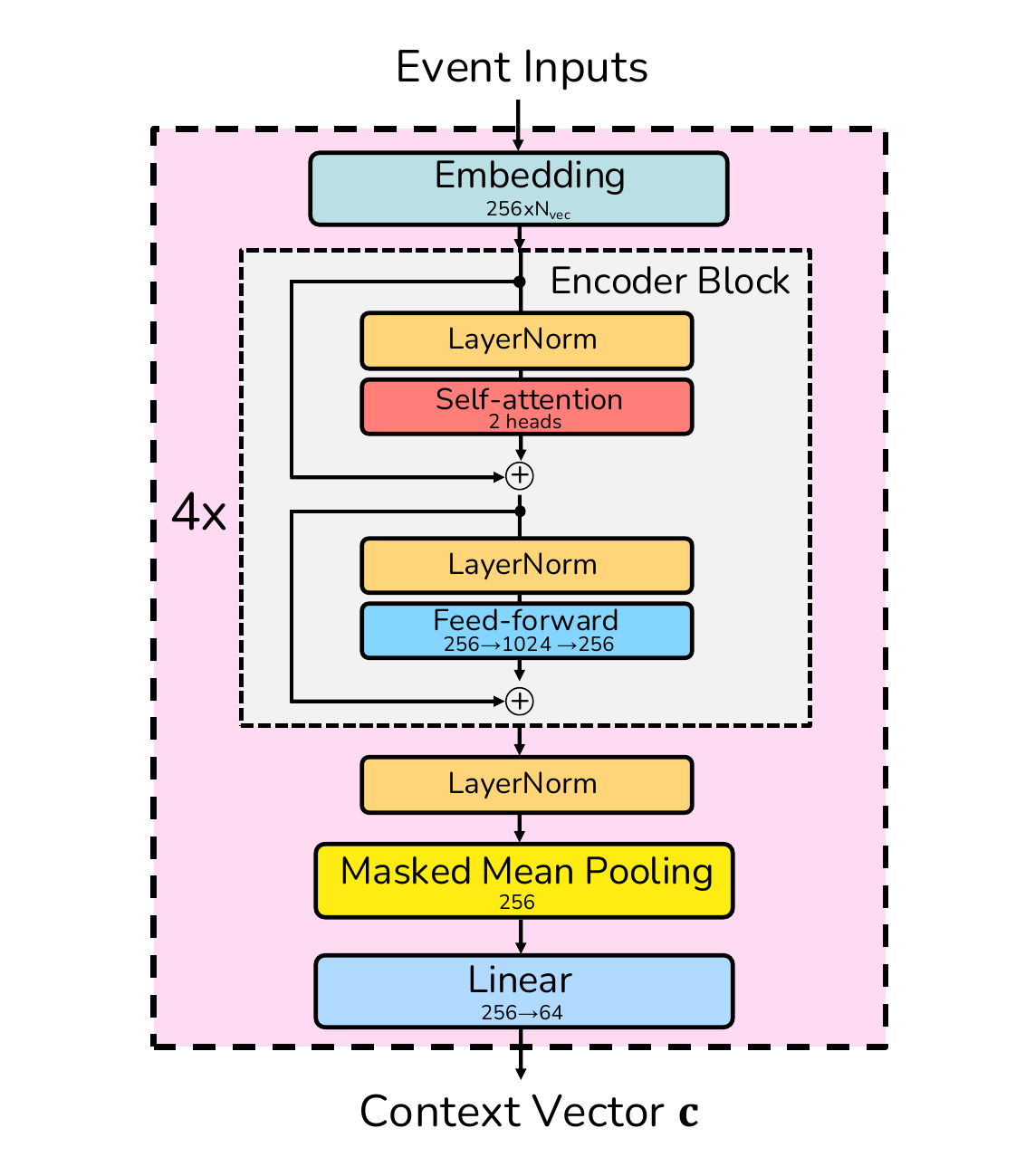}
\caption{Diagram of the transformer conditioning network, used to determine the context vector $\mathbf{c}$, a learned summary of the event. The decay mode flags are used to determine which four-vectors and vertex vectors should be considered for a given event. The input vectors are embedded, then processed by four transformer encoder layers, before masked mean pooling and a linear layer are used to yield the 64-dimensional $\mathbf{c}$ vector.}
\label{fig:transf}
\end{figure}

The context vector $\mathbf{c}$ is determined from reconstructed event quantities using a transformer encoder conditioning network~\cite{attention}, shown in Fig.~\ref{fig:transf}. The decay mode flags are used to determine which decay product and vertex vectors are included for that event, with all others masked (\eg sub-leading pions $\pi^\pm_2/\pi^\pm_3$ are not included for single prong decays). There are up to 13 (9) vectors per event in the \tauhtauh (\taultauh) channel, which are then embedded into a 256-dimensional token space. The resulting sequence is processed by four transformer encoder layers with pre-normalization~\cite{prenorm}, two self-attention heads, and a feed-forward width of 1024. The encoded tokens are then passed through a final normalization layer, before being averaged over the present objects, and projected to the 64-dimensional context $\mathbf{c}$. The GELU activation function is used within the feed-forward layers in the conditioning network.

\medskip 

The conditioning network and the flow are trained simultaneously, minimizing the negative log-likelihood of the generator-level neutrino kinematics under the learned density:
\begin{equation}
  \mathcal{L}
  = -\frac{1}{N} \sum_{i=1}^{N} \log p\!\left(\bm{\nu}_i  \mid \mathbf{c}_i\right),
  \label{eq:nll}
\end{equation}
where $\bm{\nu}_i$ denotes the true neutrino kinematics of event $i$, $\mathbf{c}_i$ the context vector of the corresponding reconstructed event, and the density is evaluated via Eq.~(\ref{eq:changevars}).

Optimization is performed with the \textsc{AdamW} optimizer~\cite{adamw} at a base learning rate of $5\times10^{-4}$ and batch size of 4096, with a linear warm-up over the first 5\% of training steps, followed by a cosine decay to 1\% of the base rate. The model is trained for 100 epochs, and the checkpoint with the lowest validation loss is retained.

\medskip

At inference, the context vector $\mathbf{c}$ is computed, and the predicted conditional density is sampled by drawing a latent vector from $\mathcal{N}(\mathbf{z};\mathbf{0},\mathbb{I})$ and applying the inverse transformation:
\begin{equation}
    \bm{\nu}_{pred} =  f^{-1}(\mathbf{z} \mid \mathbf{c})
 \end{equation}
The distribution of the predicted density can be determined by sampling  multiple times in the way described. 

Alternatively, a single point can be taken as the most probable configuration, the maximum-density point (MAP).
The location of the maximum of a density (probability per unit volume), is not invariant under a change of variables~\cite{MacKay1998}, and a choice of volume element
must therefore be made.
The flow models the density of the neutrino momentum components along $\{\bm{n}_i, \bm{r}_i, \bm{k}_i\}$, with volume element
$\mathrm{d}^3p_{\nu} = |\vec{p}_\nu|^2\,\mathrm{d}|\vec{p}_\nu|\,\mathrm{d}\Omega$, where
$\mathrm{d}\Omega$ is the solid-angle element of the neutrino direction.
Since the neutrino energy spectrum in $\tau$ decays extends down to zero, this density
rises as $|\vec{p}_\nu|^{-2}$ when $|\vec{p}_\nu| \to 0$, even though little probability
lies there.
To avoid this, and obtain a prediction more useful for spin studies, we determine the MAP
with respect to the volume element $\mathrm{d}|\vec{p}_\nu|\,\mathrm{d}\Omega$ for each
neutrino system instead, which removes the $|\vec{p}_\nu|^2$ factor. The quantity
maximized is the transformed density:
\begin{equation} \label{eq:transf_density}
  \tilde{p}(\bm{\nu} \mid \mathbf{c}) = p(\bm{\nu} \mid \mathbf{c}) \prod_i |\vec{p}_{\nu_i}|^2 ,
\end{equation}
where $i$ runs over the neutrino systems.
Writing $\bm{\nu}(\mathbf{z}) = f^{-1}(\mathbf{z} \mid \mathbf{c})$, the objective follows
from Eqs.~(\ref{eq:changevars}) and~(\ref{eq:transf_density}) as:
\begin{equation}
  \begin{split}
    \log \tilde{p}\bigl(\bm{\nu}(\mathbf{z}) \mid \mathbf{c}\bigr)
    = {} & \log \mathcal{N}\!\bigl(\mathbf{z};\, \mathbf{0}, \mathbb{I}\bigr)
           - \log \left| \det J_{f^{-1}}(\mathbf{z} \mid \mathbf{c}) \right| \\
         & + 2\sum_i \log |\vec{p}_{\nu_i}(\mathbf{z})| ,
  \end{split}
  \label{eq:map}
\end{equation}
where the Jacobian of the inverse transformation satisfies
$\log \left| \det J_{f^{-1}} \right| = -\log\left| \det J_{f} \right|$.
This objective is maximized by gradient ascent on $\mathbf{z}$, with the network parameters held fixed and an initial guess of $\mathbf{z}=\mathbf{0}$.
The MAP neutrino estimate is then $\hat{\bm{\nu}}_{pred} = f^{-1}(\hat{\mathbf{z}} \mid \mathbf{c})$.
The optimization is performed in the latent space to exploit the isotropy of the Gaussian base distribution, which contributes a well-conditioned quadratic term to Eq.~(\ref{eq:map}) and stabilizes the gradient ascent.
This approach was found to converge more reliably than maximizing directly over the neutrino components.
The \textsc{Adam} optimizer~\cite{adam} was used for this task, with 200 steps and a learning rate of $10^{-2}$.

\subsection{Comparison with Transformer Architecture}

The conditional flow architecture is compared with a stand-alone transformer setup. This architecture is derived from the transformer conditioning network in Fig.~\ref{fig:transf}, with a regression head added. The regression head consists of a $64\to64$ dimensional linear layer, followed by a GELU layer and a final $64\to D$ linear layer.

The optimization is performed using the \textsc{AdamW} optimizer for 100 epochs, with the checkpoint achieving the lowest validation loss retained. The loss is the mean-squared error between the predicted and generator-level neutrino components:
\begin{equation}
  \mathcal{L}_{\mathrm{MSE}}
  = \frac{1}{N} \sum_{i=1}^{N} \bigl\| {\bm{\nu}_{pred}}_{i} - \bm{\nu}_i \bigr\|^2 ,
  \label{eq:mse}
\end{equation}
where $\bm{\nu}_{pred}$ are the predicted neutrino kinematics, and $\bm{\nu}_i$ are the true neutrino kinematics, for event $i$.

\subsection{Neutrino Regression Performance}

The trained models are evaluated on an independent test sample of 3  (2.5) million Higgs events in the \tauhtauh (\taultauh) channel. Two inference modes of the \texttt{TauPolaris} conditional flow are considered (sampled and MAP), and are compared throughout with a stand-alone transformer regressor trained on the same inputs.
The capacity of the models to learn the generator distributions is evaluated by comparing the predicted and generator-level distributions of kinematic quantities. 

\begin{figure*}[!htbp]
\includegraphics[width=0.32\linewidth]{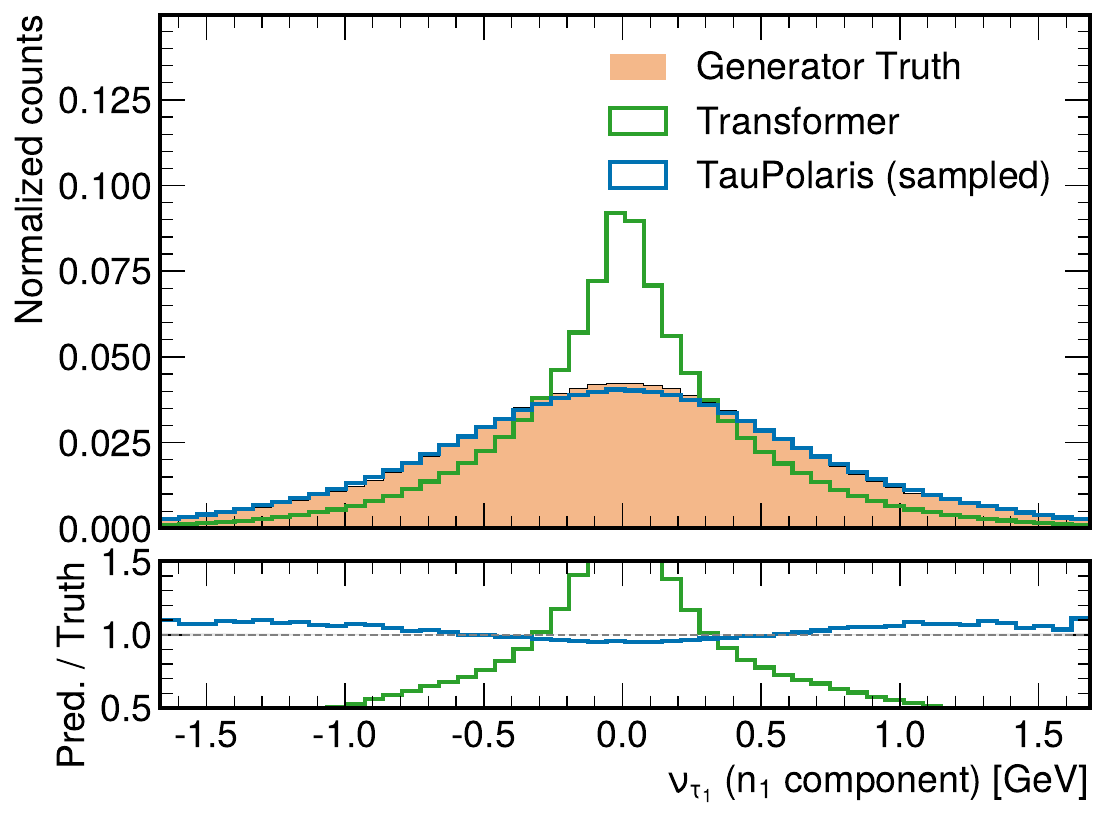}
\includegraphics[width=0.31\linewidth]{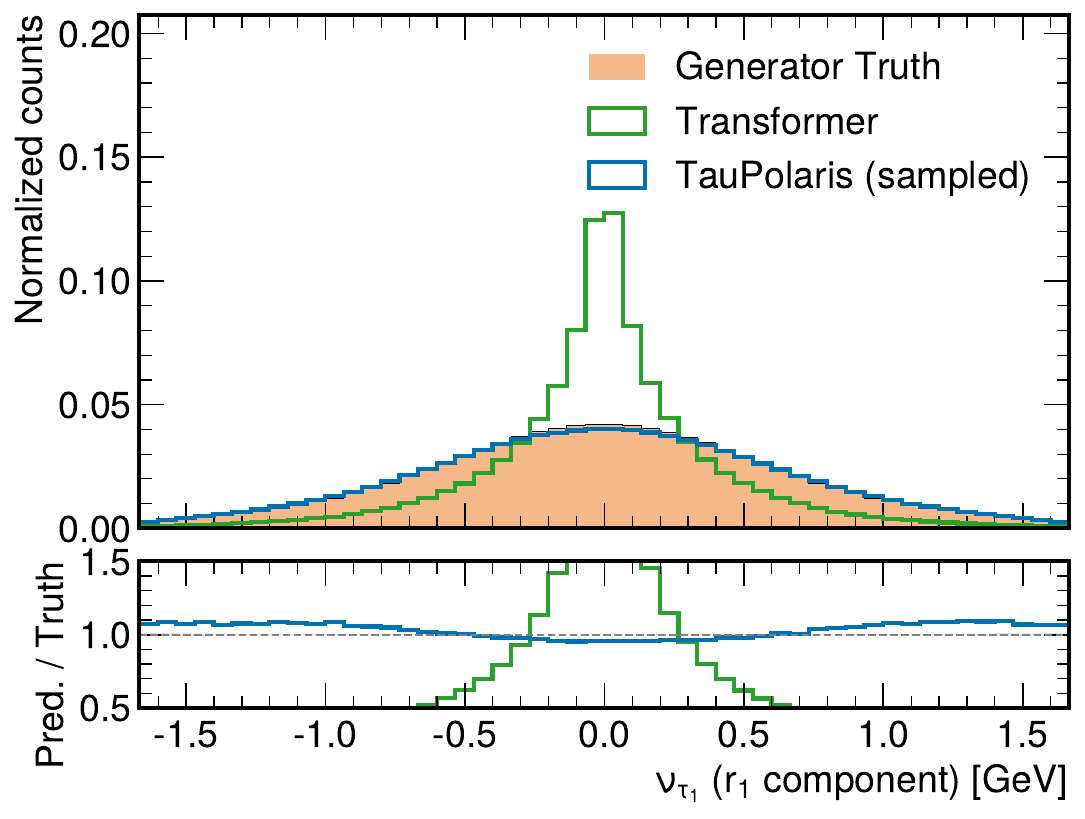}
\includegraphics[width=0.31\linewidth]{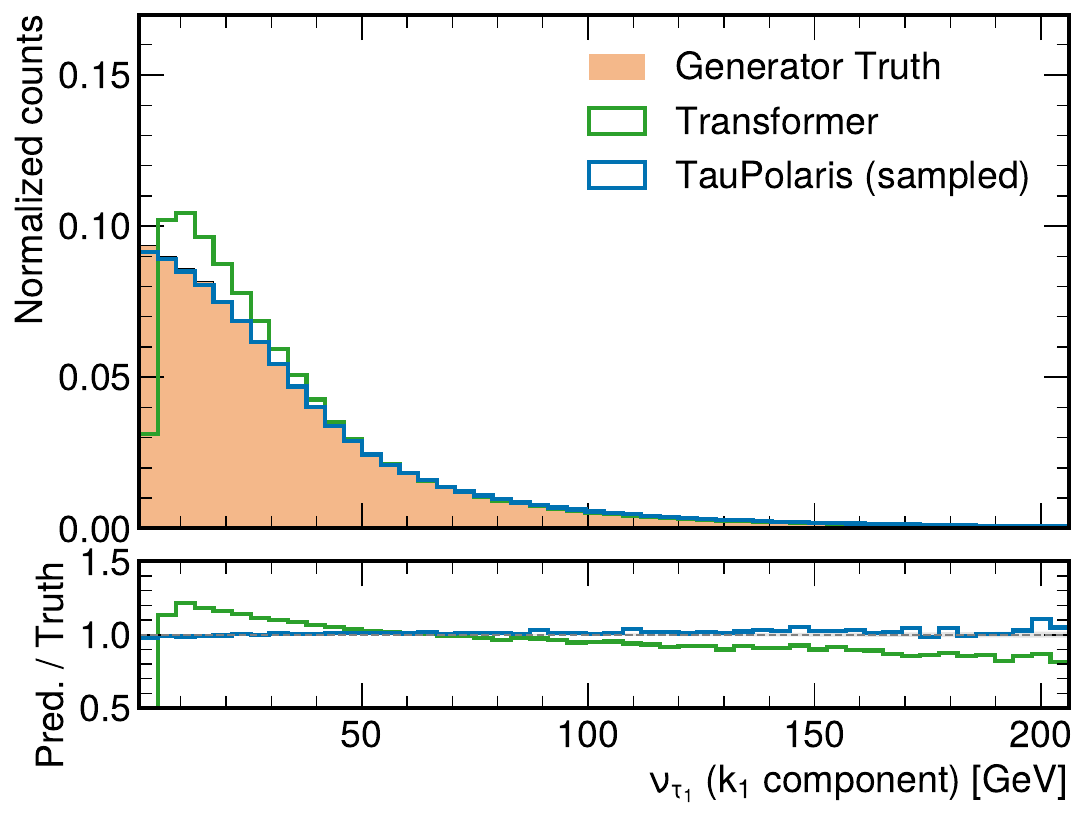}
\caption{Distributions of the regression targets for $\bm{\nu_{\tau_1}}$ in the \tauhtauh channel, comparing predictions from the \texttt{TauPolaris} and stand-alone transformer models to the generator truth.}
\label{fig:kinematics}
\end{figure*}

The predicted neutrino components are displayed in Fig.~\ref{fig:kinematics} for $\bm{\nu_{\tau_1}}$ in the \tauhtauh final states, which represent a larger fraction of ditau events than the \taultauh channel, and typically have higher sensitivity.
Similar agreement is observed for $\bm{\nu_{\tau_2}}$, and in the \taultauh channel.
Sampling from the \texttt{TauPolaris} predicted density reproduces the generator-level distributions closely across all components.
The stand-alone transformer, which predicts the same components under a mean-squared-error objective, reproduces the $k_1$ component (in the direction of the visible $\tau$) with reasonable accuracy (right), but does not reproduce the expected distributions for the $n_1$ and $r_1$ components (in the transverse plane). 
This reflects a tendency of the transformer to predict a neutrino that is collinear with the visible $\tau$. 
The accurate modeling of the neutrino direction is crucial for spin measurements, and while individual components may be well modeled, it is important to ensure that correlations between them are respected.

\begin{figure*}[!htbp]
\includegraphics[width=0.32\linewidth]{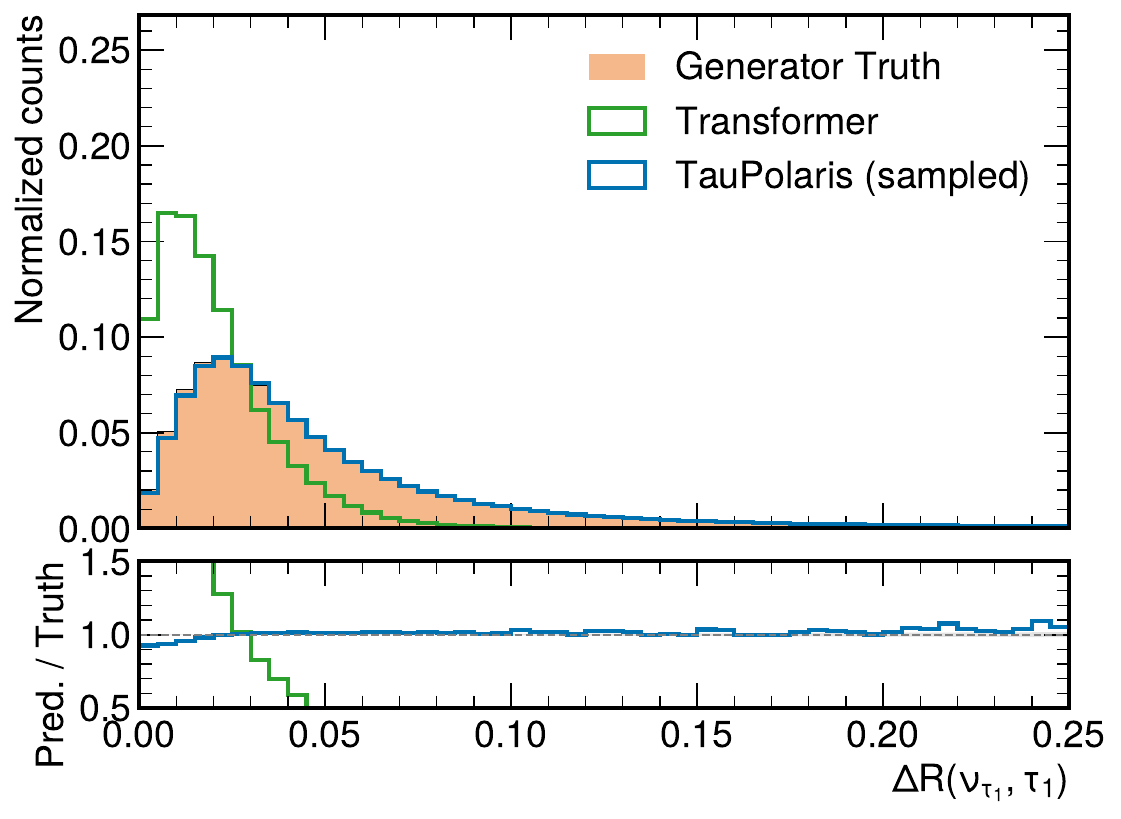}
\includegraphics[width=0.31\linewidth]{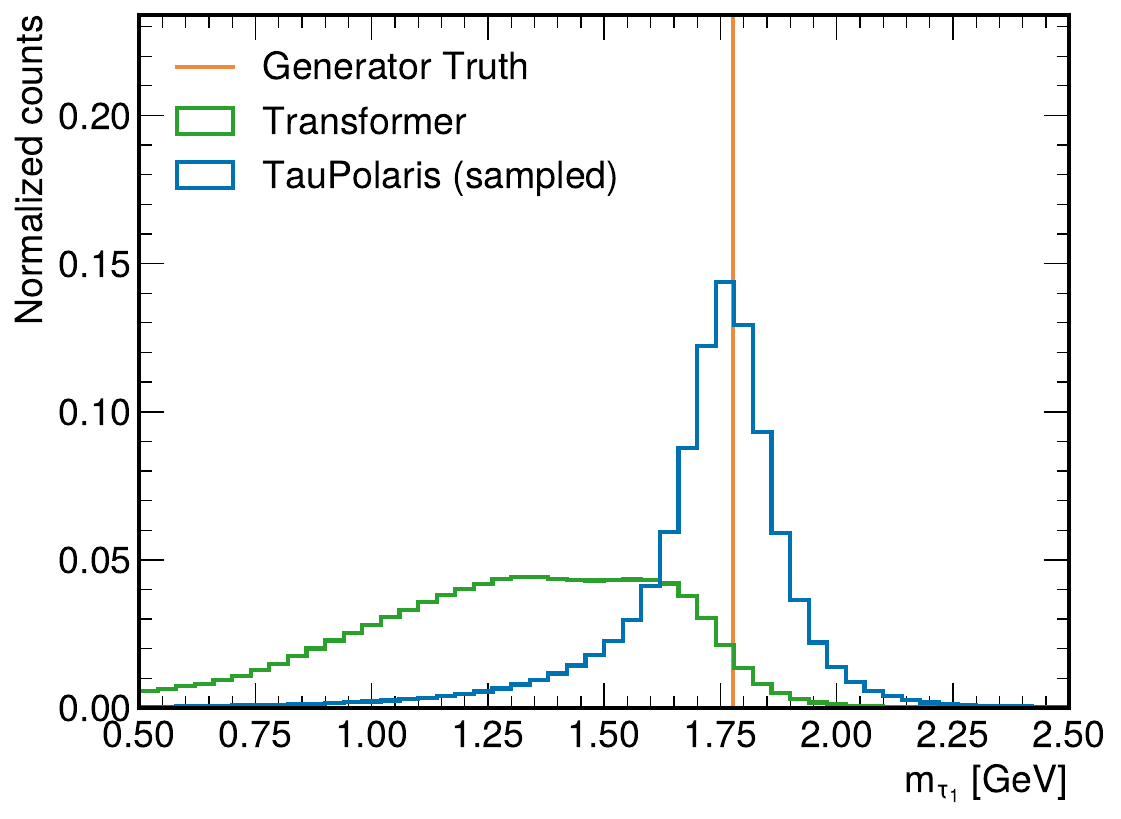}
\includegraphics[width=0.31\linewidth]{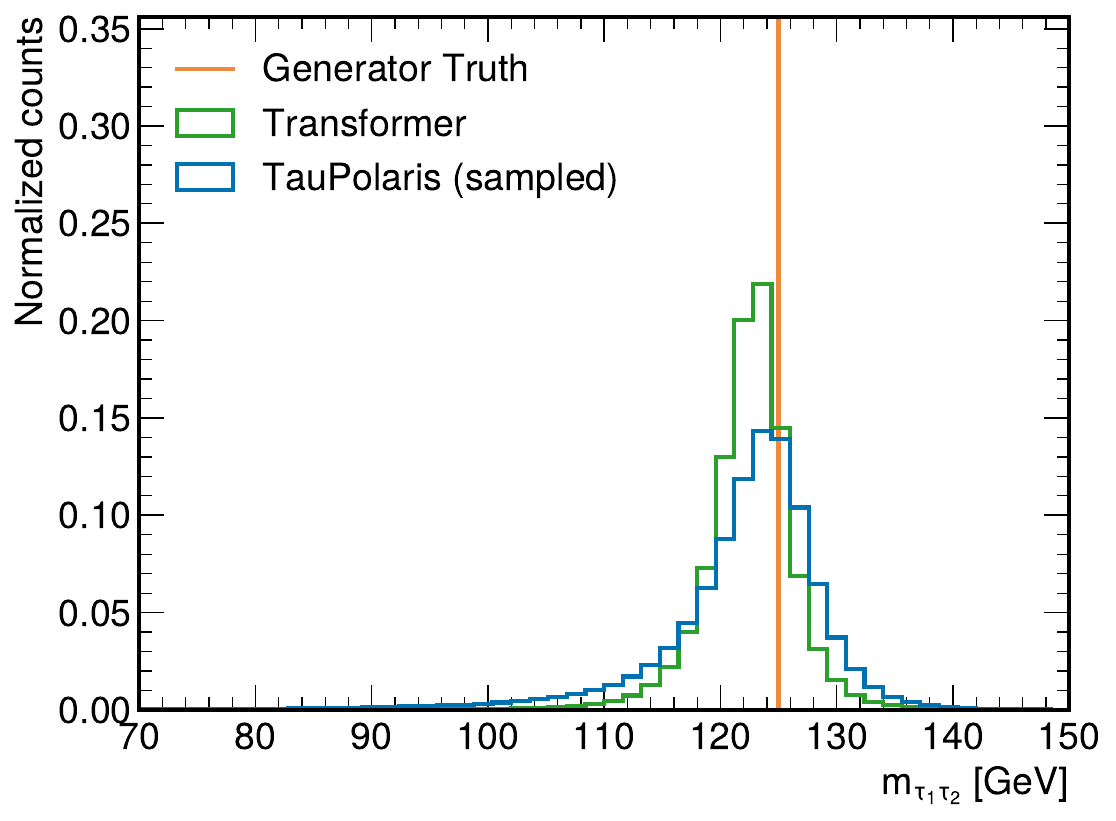}
\caption{Distributions of variables which depend on correlations between momentum components, comparing predictions from the \texttt{TauPolaris} and stand-alone transformer models to the generator truth in the \tauhtauh channel. The full $\tau$ lepton momentum is obtained by summing the visible decay products with the estimated neutrino. The angular separation between the neutrino and full $\tau$ momentum is shown on the left, and the invariant mass of the $\tau$ lepton is shown in the center. Similar agreement is seen for the other $\tau$ lepton in the pair. The combined ditau invariant mass is shown on the right.}
\label{fig:correlated_kinematics}
\end{figure*}

Distributions of some variables built by combining neutrino momentum components are shown in Fig.~\ref{fig:correlated_kinematics} for the \tauhtauh channel.
The full $\tau$ lepton momenta are obtained by combining the predicted neutrinos with the reconstructed visible momenta.
The angular separation between the neutrino and the $\tau$ lepton momentum is shown on the left. This is well modeled by \texttt{TauPolaris}, and further illustrates the tendency of the transformer to predict a neutrino that is nearly collinear with the visible $\tau$, as evidenced by the excess of events near $\Delta R(\nu_{\tau_1}, \tau_1) \approx 0$.
The invariant mass of the $\tau$ lepton is shown in the middle. This is another quantity which is difficult to regress, as it requires accurate modeling of the correlations between different components of the neutrino momenta.
The \texttt{TauPolaris} predicted density is correctly centered on the physical value ($m_\tau = 1.78$~GeV). This is a consequence of the flow modeling the full joint density of the neutrino components rather than each component independently. The transformer struggles to preserve these correlations, and its reconstructed $\tau$ mass is consequently biased low, since independent per-component errors do not respect the mass constraint.
The invariant mass of the ditau system is shown on the right. Both \texttt{TauPolaris} and the transformer predict a distribution close to the expected value. In this case, the transformer prediction is more narrowly distributed, but as illustrated by the previous plots, this does not lead to better performance for spin-related measurements.

\begin{figure}[!htbp]
\includegraphics[width=\linewidth]{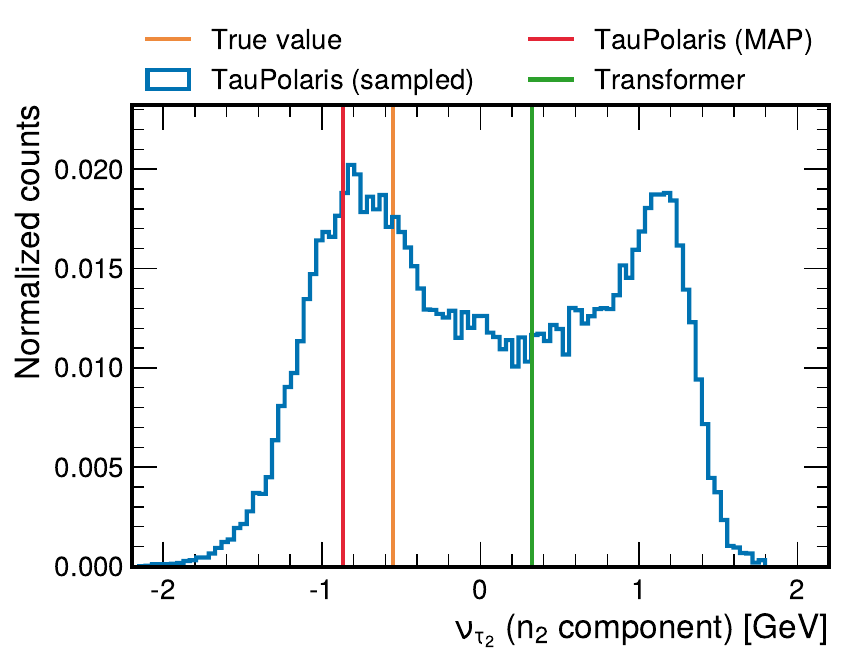} \\ 
\includegraphics[width=\linewidth]{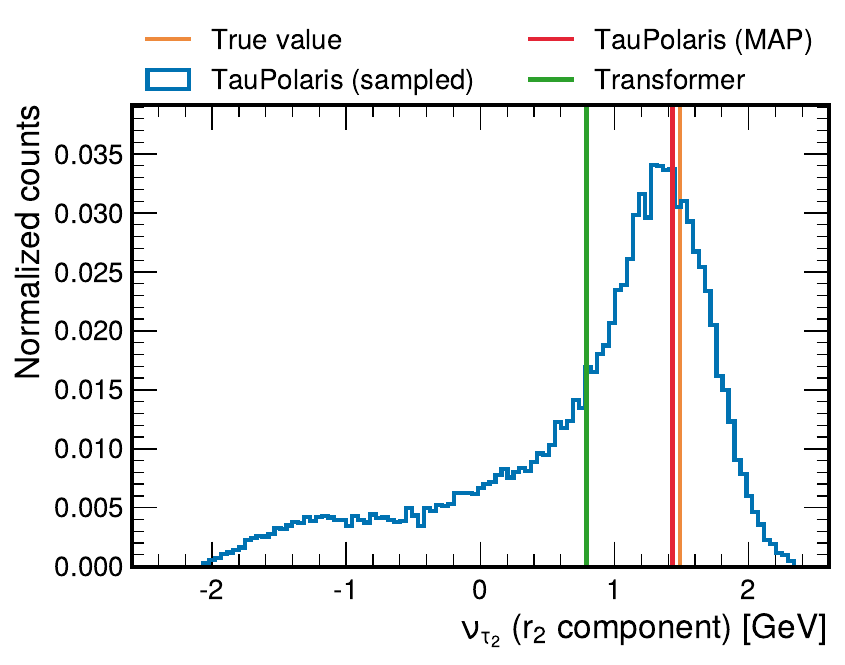}
\caption{Predicted neutrino components orthogonal to the visible $\tau$ lepton direction for two different types of event in the \tauhtauh channel. The generator-level truth is compared to sampled and MAP predictions from \texttt{TauPolaris}, as well as the stand-alone transformer for comparison. The MAP can provide a more accurate prediction than a single random sample by selecting the most probable neutrino momentum value. A multimodal sampled distribution is shown in the upper plot, while the lower plot shows an event with an asymmetric tail. }
\label{fig:pdfs}
\end{figure}

The reason why \texttt{TauPolaris} can reproduce the expected distributions of the neutrino $n_i$ and $r_i$ momentum components, while the transformer cannot, can be understood from the learned conditional densities.
Figure~\ref{fig:pdfs} shows two examples, which are representative of the two common types of events that are seen.
The first type of event is in the upper plot, where the conditional density exhibits multimodal behavior.  
While the sign of the $k$ component is typically well constrained, the orthogonal $n_i$ and $r_i$ components can often take two possible opposite-signed values, while satisfying the kinematic constraints of the event. 
The \texttt{TauPolaris} sampled distribution reveals this bimodal density. The MAP estimate from the flow can be used to preferentially select the most probable value of the neutrino momentum, which will typically lie closer to the generator-level truth than a single sample from the density. 
The stand-alone transformer prediction is also overlaid. Because of the degeneracy in these cases, the transformer typically regresses toward the mean value, which is close to zero because the peaks have opposite sign.
This contributes to the collinearity discussed above, and reduces its ability to correctly describe spin correlations.
The second type of event is in the lower plot, where there is instead a long asymmetric tail. In this case, the stand-alone transformer prediction can be closer to the maximum, but may still be biased by the tail, which typically extends past zero to the opposite sign of the main peak. This leads to predictions for the transverse components which are closer to zero. 
The \texttt{TauPolaris} prediction benefits from the knowledge of the full conditional density in both cases.

By construction, the distribution of MAP estimates does not reproduce the target kinematic distribution, as the mode of a density is not itself distributed as the density.
The MAP is used only as a per-event point estimate, and assessed by its resolution. While the best overall agreement of the distributions is obtained by sampling, the MAP can provide a more accurate prediction than a single random sample.

\begin{figure*}[!htbp]
\includegraphics[width=0.31\linewidth]{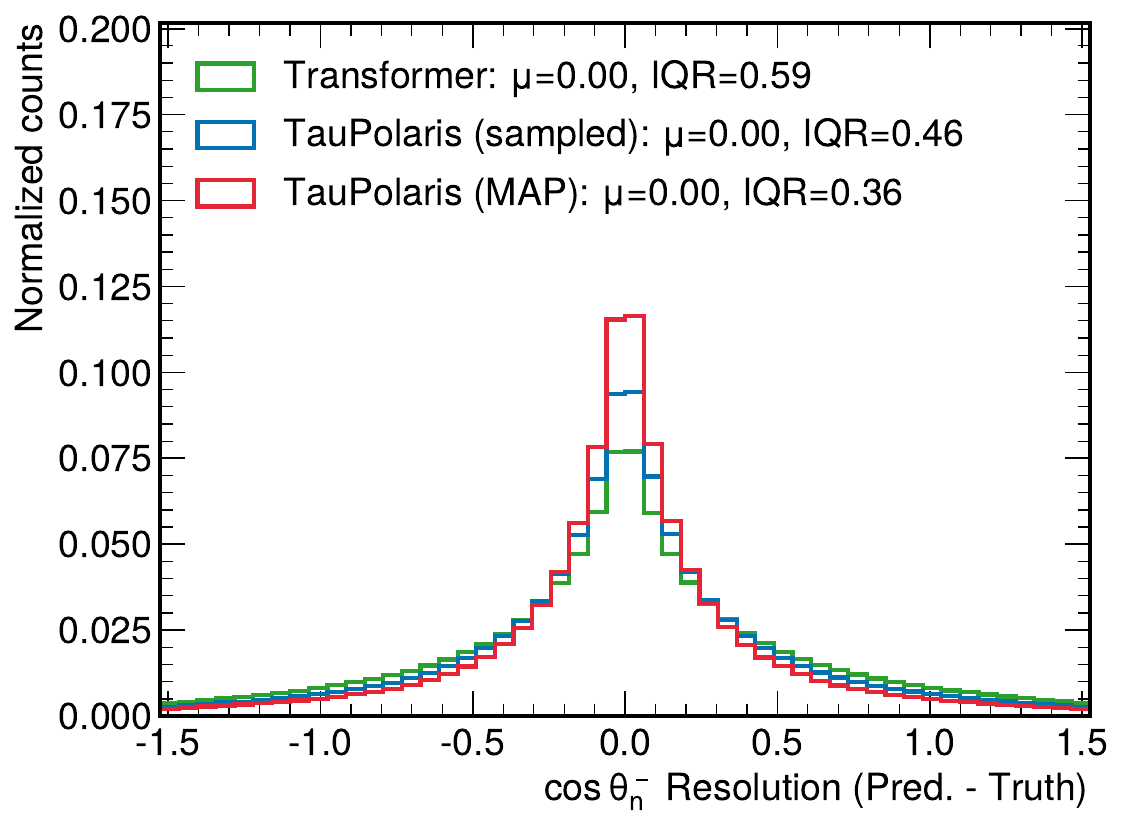}
\includegraphics[width=0.31\linewidth]{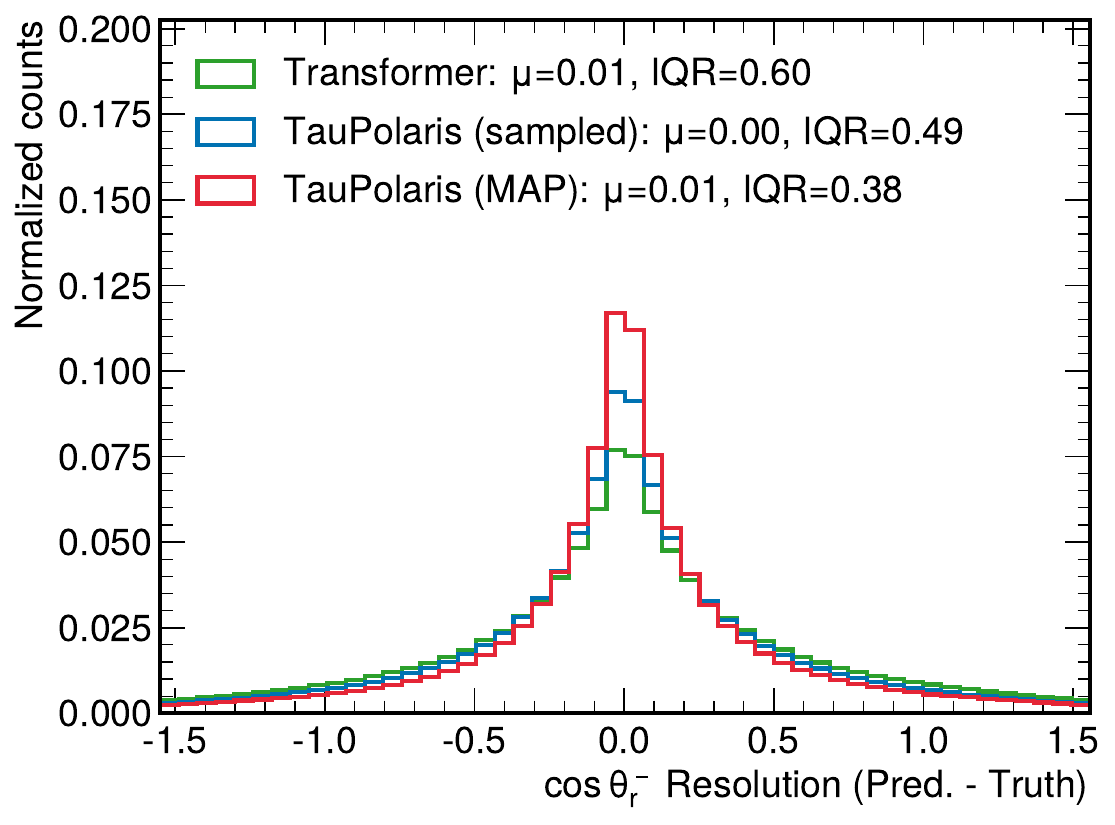}
\includegraphics[width=0.31\linewidth]{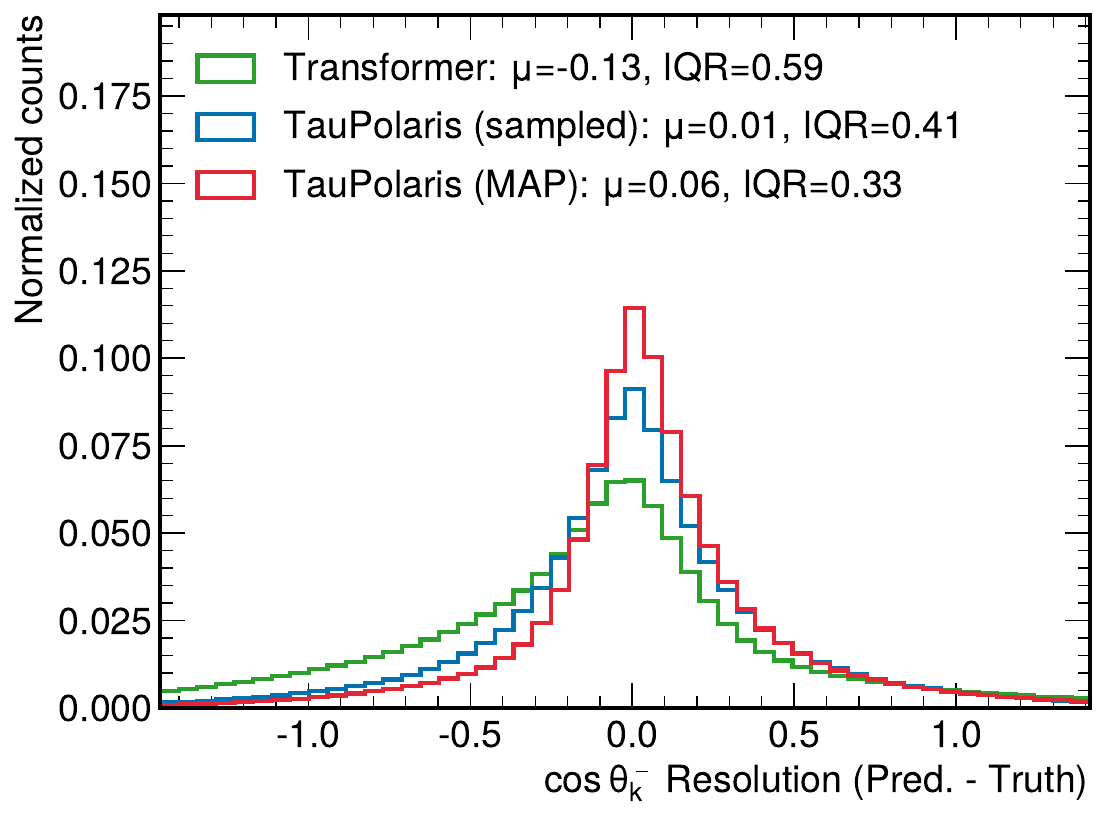}
\caption{Resolution of the \texttt{TauPolaris} algorithm (sampled and MAP) compared with the stand-alone transformer in simulated $H\to\tau\tau$ events in the \tauhtauh channel, shown for the spin variables $\cos\theta_n^-$ (left), $\cos\theta_r^-$ (middle) and $\cos\theta_k^-$ (right).
The resolution is defined as the interquartile range (IQR) of the prediction minus the generator-level truth. The \texttt{TauPolaris} MAP resolution is improved by approximately 40\% compared with the stand-alone transformer, and 20\% compared with the sampled approach. }
\label{fig:res}
\end{figure*}

The resolutions achieved for the spin-sensitive projections $\cos\theta_n^-$, $\cos\theta_r^-$, and $\cos\theta_k^-$ are shown in Fig.~\ref{fig:res} for the negatively charged $\tauh$ candidate in the \tauhtauh channel. Similar resolution is achieved for the positively charged candidate, and in the \taultauh channel.
These variables and their products are sensitive to the $B^\pm_i$ and $C_{ij}$ components of the spin density matrix, respectively.
The resolution is defined as the interquartile range (IQR) of the difference between the prediction and the generator-level truth. The resolution for the \texttt{TauPolaris} MAP approach is improved by approximately 40\% compared with the stand-alone transformer, and 20\% compared with the sampled approach, where the improvement is quoted as the fractional reduction in the IQR. 

\begin{figure*}[!htbp]
\includegraphics[width=0.31\linewidth]{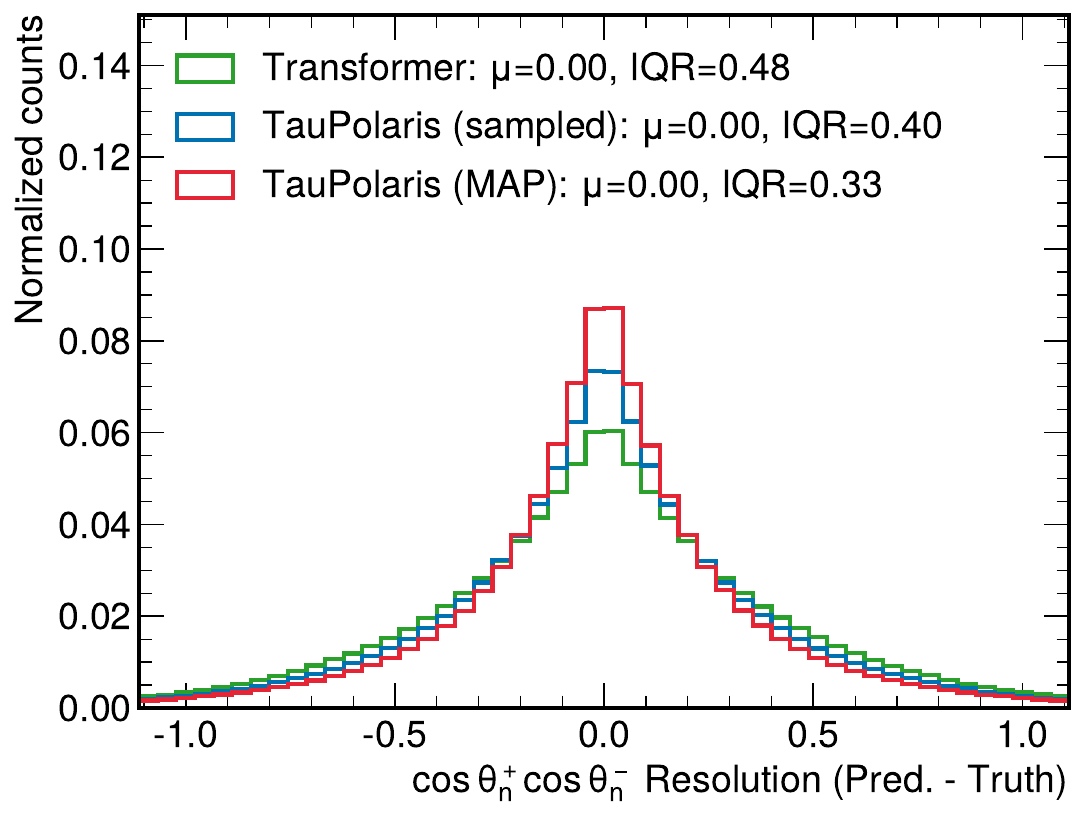}
\includegraphics[width=0.31\linewidth]{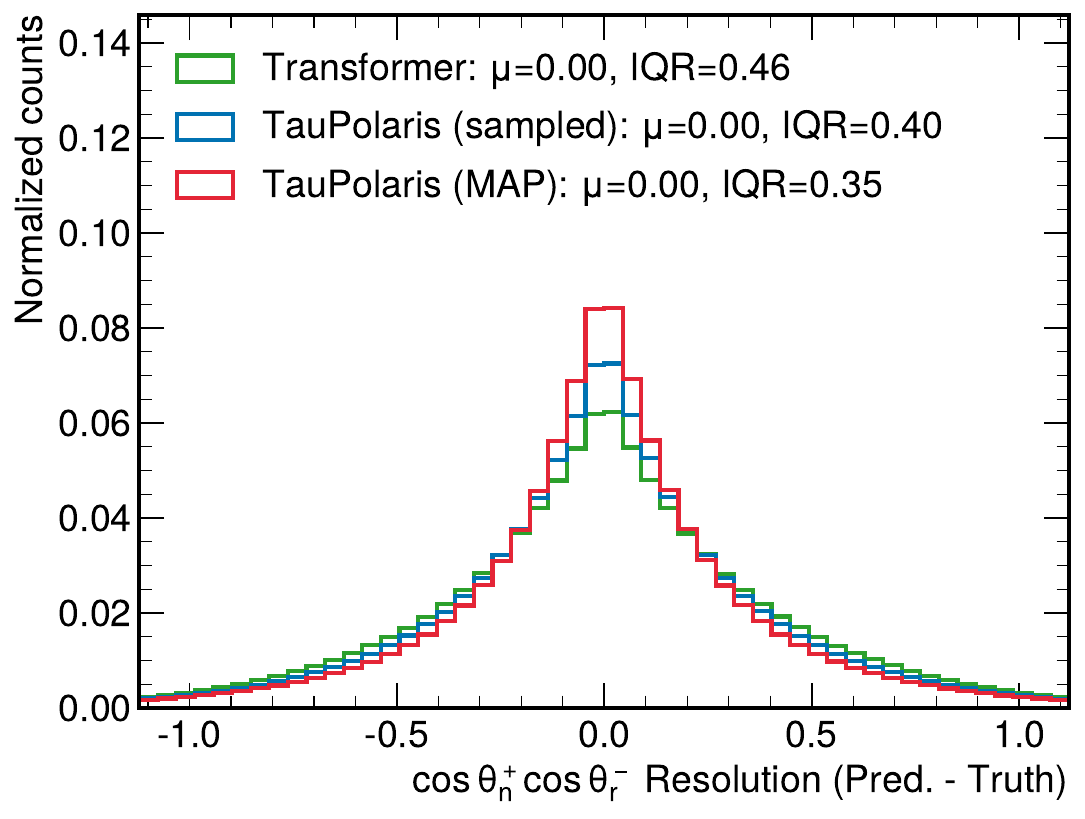}
\includegraphics[width=0.31\linewidth]{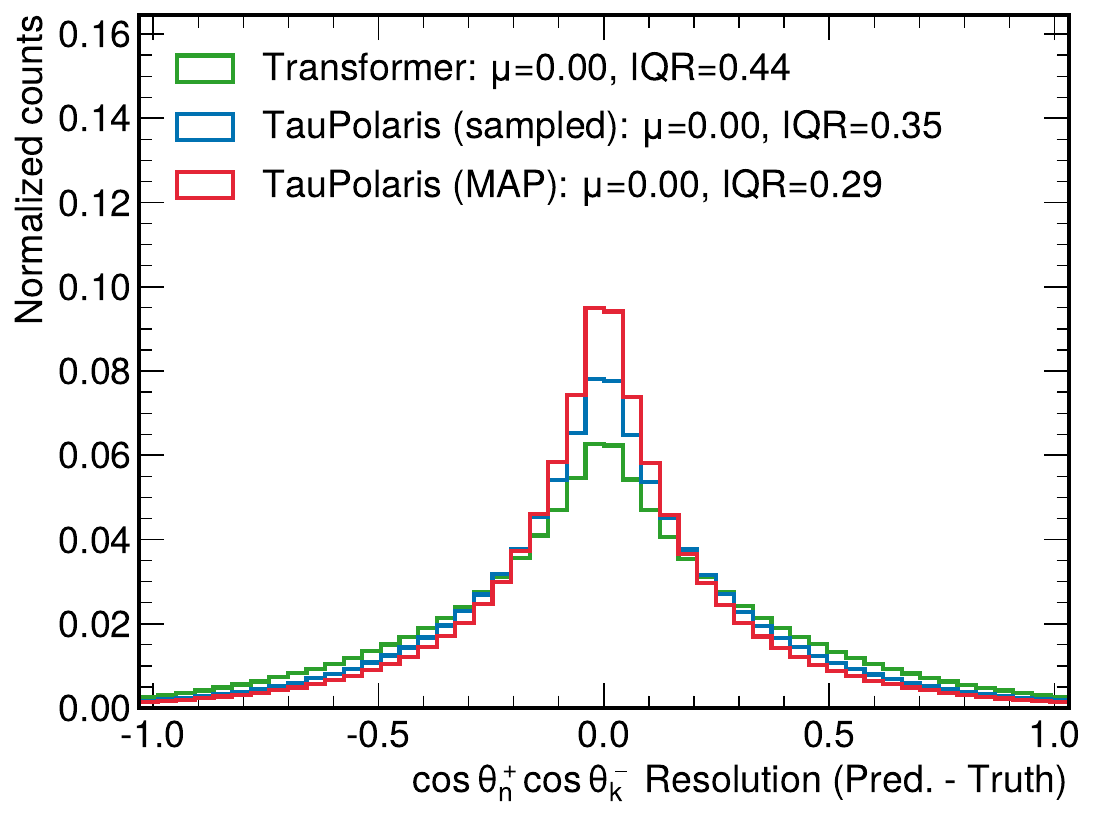}
\includegraphics[width=0.31\linewidth]{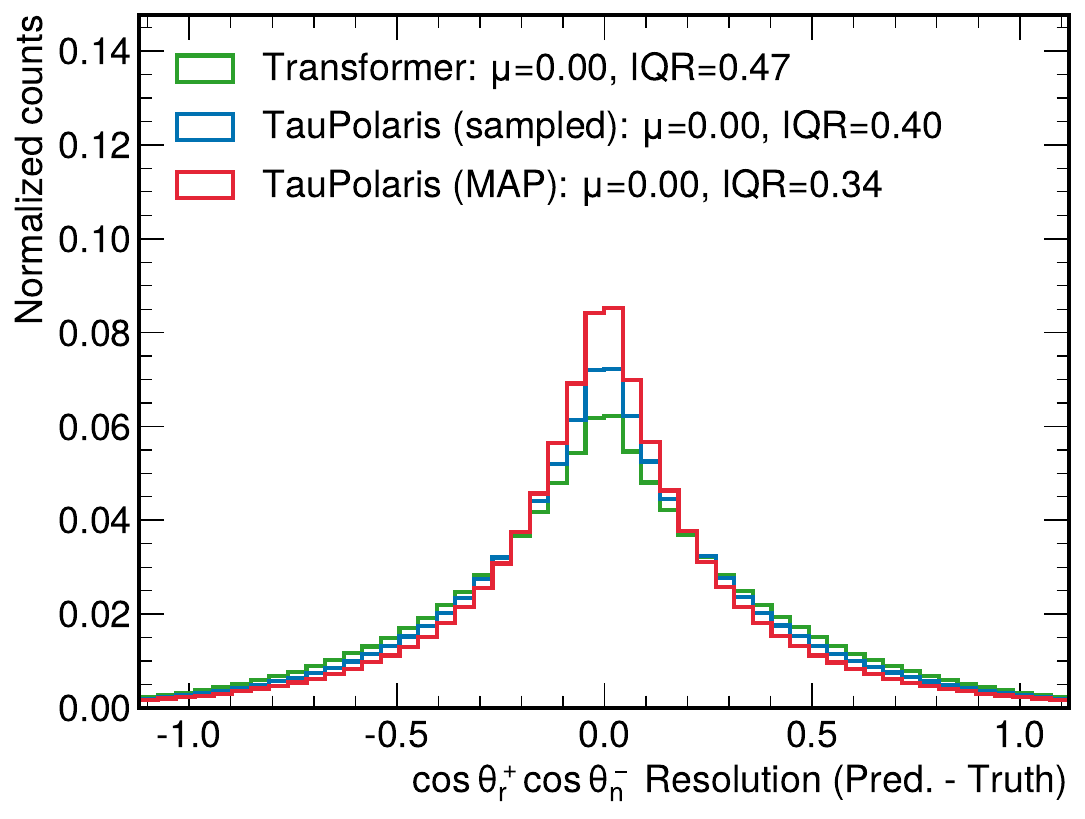}
\includegraphics[width=0.31\linewidth]{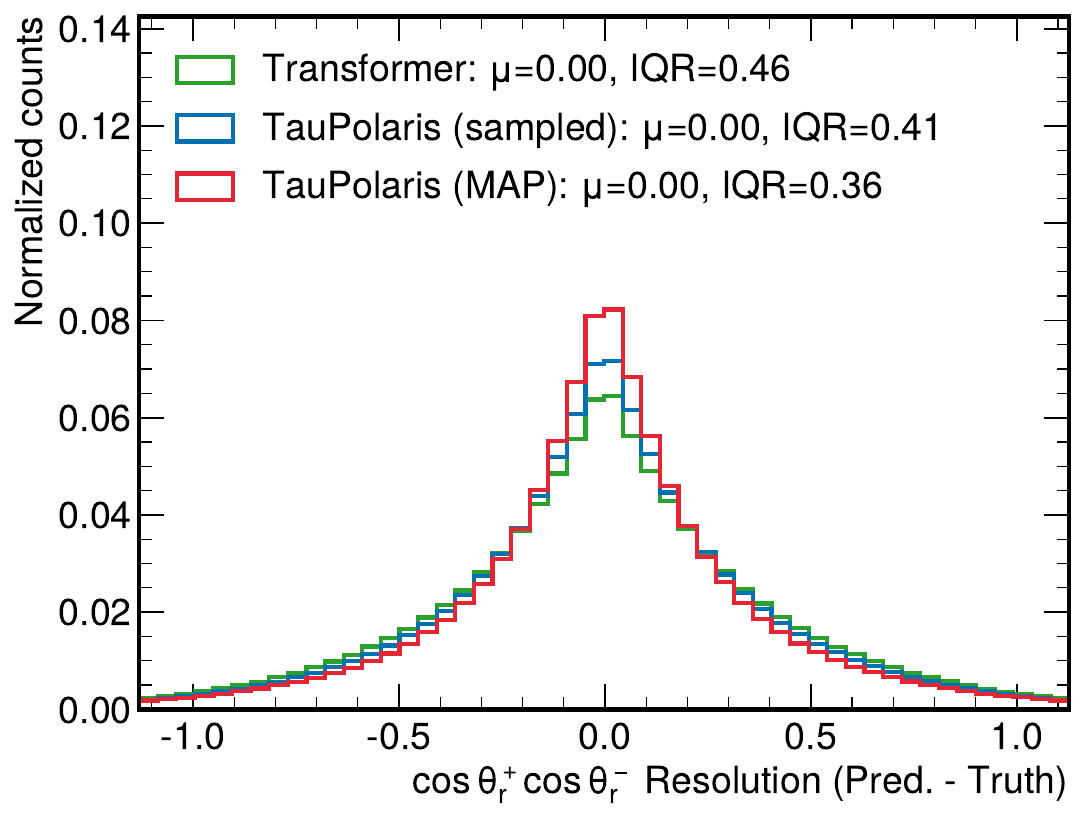}
\includegraphics[width=0.31\linewidth]{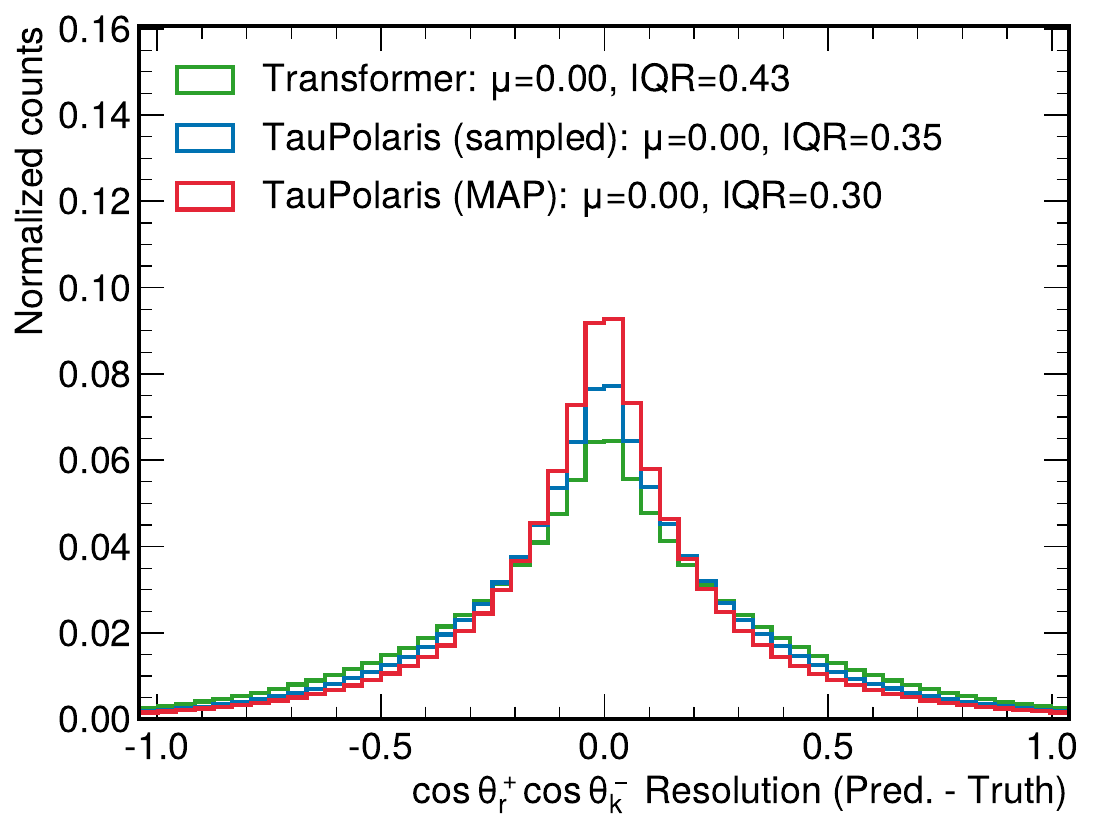}
\includegraphics[width=0.31\linewidth]{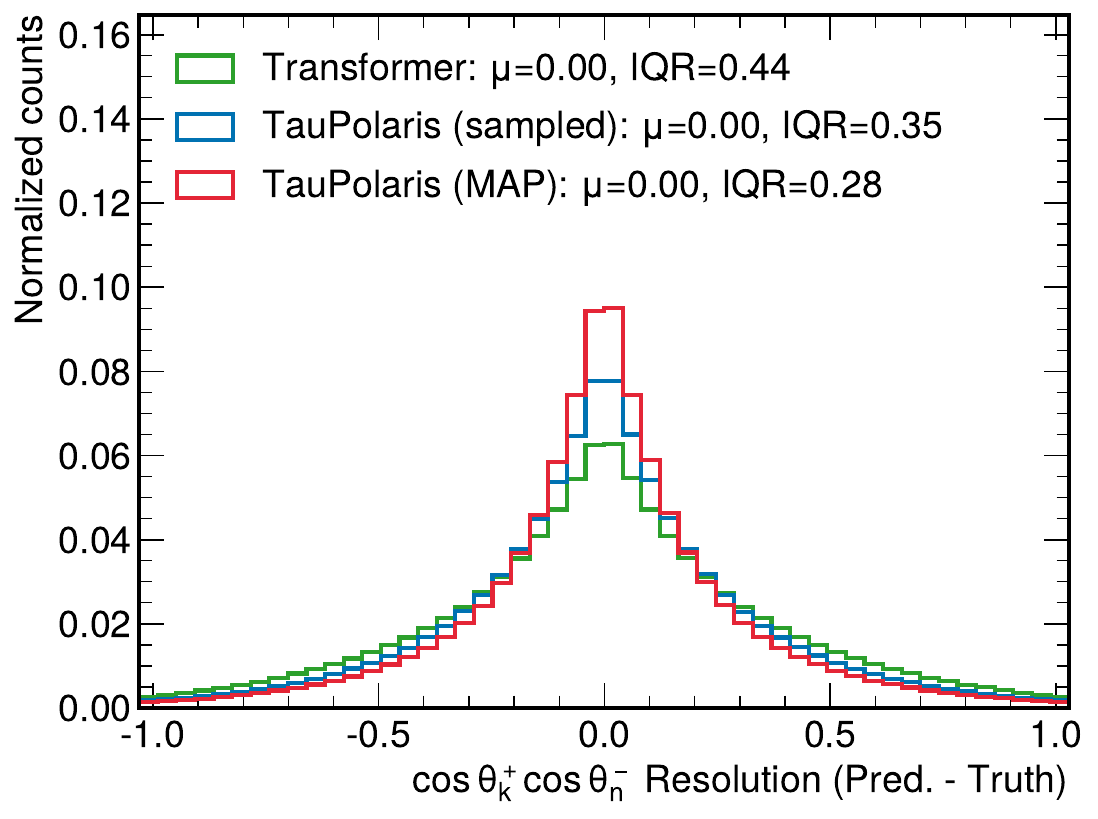}
\includegraphics[width=0.31\linewidth]{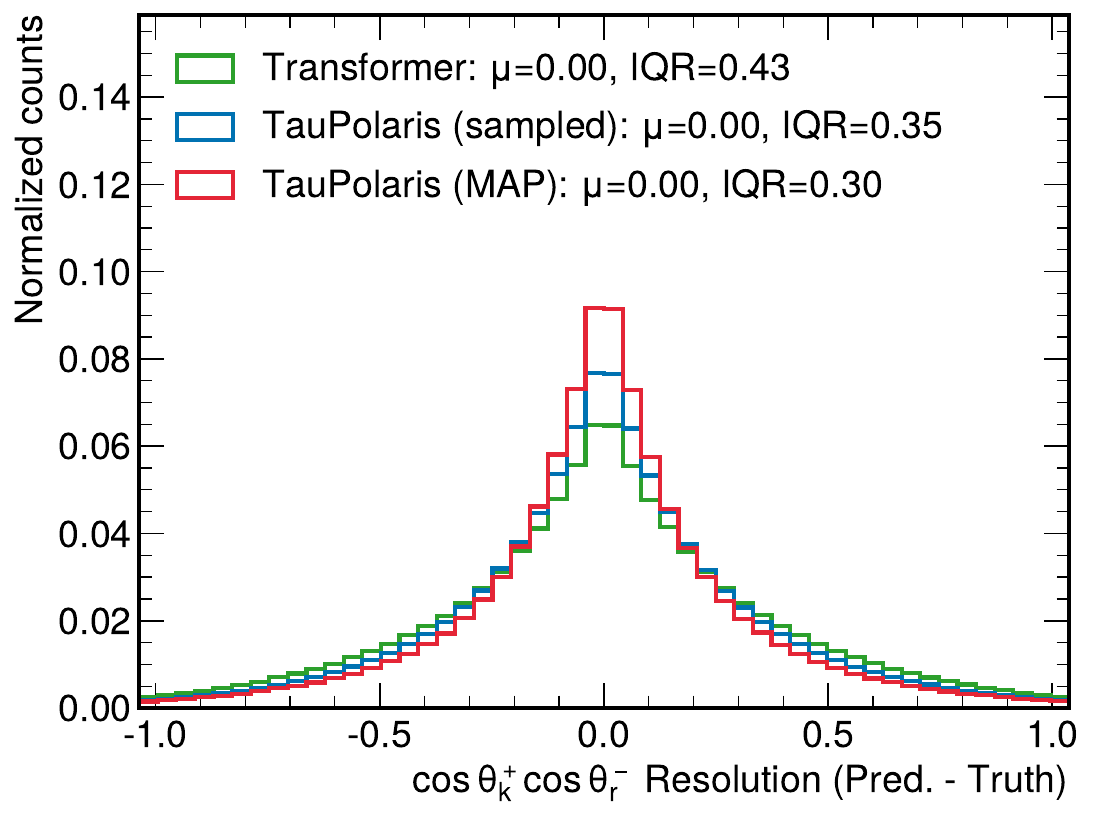}
\includegraphics[width=0.31\linewidth]{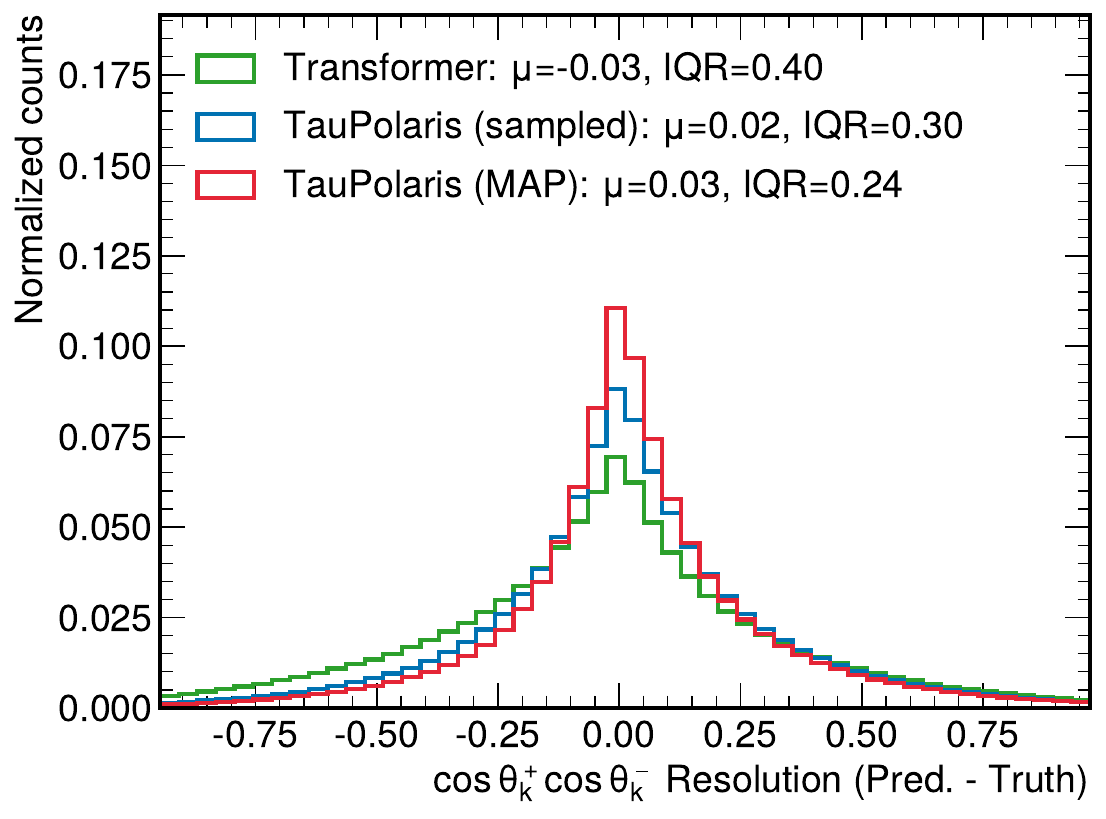}
\caption{Resolution of the \texttt{TauPolaris} algorithm (sampled and MAP) compared with the stand-alone transformer in simulated $H\to\tau\tau$ events in the \tauhtauh channel, shown for the products of the angular variables sensitive to the $C_{ij}$ components of the spin density matrix. 
The resolution is defined as the interquartile range (IQR) of the prediction minus the generator-level truth. The \texttt{TauPolaris} MAP resolution is improved by approximately 30\% compared with the stand-alone transformer, and 15\% compared with the sampled approach.}
\label{fig:res_density}
\end{figure*}

The resolutions achieved for the products of the $\cos\theta_n^\pm$, $\cos\theta_r^\pm$, and $\cos\theta_k^\pm$ projections, which are sensitive to the $C_{ij}$ components of the spin density matrix, are shown in Fig.~\ref{fig:res_density}. 
The average improvement in resolution with the \texttt{TauPolaris} MAP is approximately 30\% compared with the stand-alone transformer and 15\% compared with the sampled approach.

\subsection{Determining polarimetric vectors}

The estimated neutrino is used, along with the visible $\tau$ decay products, to determine the polarimetric vector. The procedure depends on the $\tau$ decay topology. The $\tau$ lepton momentum, and therefore the $\tau$ lepton rest-frame, is determined by combining the visible decay product momenta with the regressed neutrino.

The polarimetric vector for the $1\pi^\pm0\pi^0$ decay mode is simply the spatial part of the charged pion four-momentum, $\boldsymbol{p}_\pi$ in the $\tau$ lepton rest-frame~\cite{cherepanov2018polvec}.

The polarimetric vector for the $1\pi^\pm1\pi^0$ decay is constructed in the $\tau$ lepton rest
frame from the charged pion four-momentum, $\boldsymbol{p}_\pi$, the neutral pion four-momentum, $\boldsymbol{p}_{\pi^0}$, and the neutrino four-momentum, $\boldsymbol{p}_\nu$. Defining
$\boldsymbol{q} = \boldsymbol{p}_\pi - \boldsymbol{p}_{\pi^0}$, the polarimetric vector
is given by the spatial part of
\begin{equation}\label{eq:pvec_dm1}
\boldsymbol{h} = \frac{m_\tau\big[2(\boldsymbol{q}\cdot \boldsymbol{p}_\nu)\,\boldsymbol{q} - 
\boldsymbol{q}^2\,\boldsymbol{p}_\nu\big]}
{2(\boldsymbol{q}\cdot \boldsymbol{p}_\nu)(\boldsymbol{q}\cdot \boldsymbol{p}_\tau) - 
\boldsymbol{q}^2(\boldsymbol{p}_\nu\cdot \boldsymbol{p}_\tau)},
\end{equation}
where $m_\tau$ is the mass of the $\tau$ lepton~\cite{cherepanov2018polvec}. 

The same formalism is employed for $1\pi^\pm2\pi^0$ decays as an approximation, since the exact formula cannot be used because the two $\pi^0$s cannot be resolved. The polarimetric vector is computed using Eq.~(\ref{eq:pvec_dm1}), with $\boldsymbol{p}_{\pi^0}$ taken as the sum of both neutral pion four-momenta. The $\boldsymbol{h}$ vector obtained remains correlated with the true polarimetric vector. Furthermore, a large fraction of $\tau$ leptons reconstructed as $1\pi^\pm2\pi^0$ are in reality $1\pi^\pm1\pi^0$, as a consequence of the larger branching fraction of the latter. 

The polarimetric vector for $3\pi^\pm0\pi^0$ decays is defined using the $\tau$ lepton and visible $\tau$ momenta as described in Ref.~\cite{Cherepanov:2023wfp}, using the $\textrm{a}_1$ resonance model with the parametrization from Ref.~\cite{Asner:1999kj}.

The \texttt{TAUOLA} program is used to determine the polarimetric vector for $3\pi^\pm1\pi^0$ decays directly using the $\tau$ lepton decay products, including the regressed neutrino.

The polarimetric vector for $\taul$ decays is the negative of the spatial part of the light lepton momentum in the $\tau$ lepton rest frame. The presence of two neutrinos in the decay prevents the usage of the exact formula, however similarly to the $1\pi^\pm2\pi^0$ case, the $\boldsymbol{h}$ vector obtained remains correlated with the true polarimetric vector. The negative of the lepton momentum is taken because the spectral function (which encapsulates the correlation between the spin and momentum) has the opposite sign for $\taul$ decays compared with single-pion $\tauh$ decays~\cite{Berge:2011ij}.

\section{Application to Quantum Entanglement Measurements}
\label{sec:entanglement}
The elements $B^\pm_i$ of the $\mathbf{B}^\pm$ vectors can be measured by fitting the distributions of the $\cos\theta_i^{\pm}$ variables introduced in Sec.~\ref{sec:simulations}, evaluated here from the polarimetric vectors reconstructed with the method described in Sec.~\ref{sec:models}. The elements $C_{ij}$ of the $\mathbf{C}$ matrix are likewise measured from the distributions of the products $\cos\theta_i^{+}\cos\theta_j^{-}$. The concurrence $\mathcal{C}$ and the quantity $m_{12}$ then follow from the reconstructed spin density matrix using Eqs.~(\ref{eqn:concurrence}) and~(\ref{eqn:m12}).

We estimate the $\cos\theta_i^{\pm}$ and $\cos\theta_i^{+}\cos\theta_j^{-}$ variables by drawing 200 samples from the learned $p(\boldsymbol{\nu}\,|\,\mathbf{c})$, computing for each sample the polarimetric vectors and the $\{\mathbf{n},\mathbf{r},\mathbf{k}\}$ basis from the reconstructed $\tau$ lepton momenta, projecting the polarimetric vectors onto the basis axes, and taking the mean of the resulting spin variables over the samples. The products $\cos\theta_i^{+}\cos\theta_j^{-}$ are averaged directly, rather than being formed from the averaged $\cos\theta_i^{\pm}$. Since $\mathbf{p}$ is fixed, the basis is determined entirely by the reconstructed $\mathbf{k}$ axis, which has an angular resolution of about $1.8^\circ$; we checked that its impact on the measured $C_{ij}$ is negligible compared with that of the resolution on the polarimetric vectors. We find that 200 samples gives a good balance between performance and evaluation speed; increasing the number of samples by a factor of five gives a negligible performance improvement. The evaluation of the 200 samples takes about 40\,ms per event on a single CPU thread when processing events in batches of $10^4$, and below 2\,ms per event on an NVIDIA V100 GPU with batches of $10^5$ events. For comparison, the \textsc{SVFit} algorithm~\cite{Bianchini:2016yrt} used by CMS requires around 0.25\,s per event~\cite{Kalinowski:2025vci}.

We have performed a feasibility study to evaluate the sensitivity expected at the High-Luminosity LHC (HL-LHC) to the elements of $\mathbf{B}^\pm$ and $\mathbf{C}$, and to the derived quantities $\mathcal{C}$ and $m_{12}$, for the $H\rightarrow\tau\tau$ process. The study considers the Higgs boson signal together with the dominant background from $Z/\gamma^*\rightarrow\tau\tau$ events.

Events are categorized according to how the two $\tau$ leptons decay, following the categorization used for the CMS $H\rightarrow\tau\tau$ \CP measurement of Ref.~\cite{CMS:2026hvv}. Three channels are considered: the fully hadronic \tauhtauh channel and the two semileptonic $\tau_\mu\tauh$ and $\tau_e\tauh$ channels. Hadronically decaying $\tau$ leptons are classified by their reconstructed decay mode as $1\pi^\pm0\pi^0$, $1\pi^\pm1\pi^0$, $1\pi^\pm2\pi^0$, or $3\pi^\pm0\pi^0$; the $3\pi^\pm1\pi^0$ decay mode is not included. There are nine categories in the \tauhtauh channel and four in each of the semileptonic channels, seventeen in total.

The expected signal and background yields for each category are estimated from the numbers reported in the HEPData record~\cite{hepdata} of Ref.~\cite{CMS:2026hvv}. That analysis uses a multiclass boosted decision tree (BDT) to separate the Higgs boson signal from the genuine-$\tau$ and misidentified-$\tau$ backgrounds; events assigned to the Higgs category are then split by the $\tau$-pair final state and analyzed in windows of increasing BDT score, corresponding to progressively higher signal-to-background ratios. For each decay-mode category we take the yields reported in the highest BDT score window, which has the most favorable signal-to-background ratio, and scale them from the $62.4\,\text{fb}^{-1}$ recorded at $\sqrt{s}=13.6\,\text{TeV}$ to the $3\,\text{ab}^{-1}$ expected at the HL-LHC. This extrapolation assumes that the signal-to-background ratio achieved in that window is preserved.

For each element $C_{ij}$ we perform a binned maximum likelihood fit using the $\cos\theta_i^{+}\cos\theta_j^{-}$ distribution, divided into 20 bins over the range $[-1,1]$, as the discriminating variable; the $B^\pm_i$ are extracted analogously from the $\cos\theta_i^{\pm}$ distributions. The expected signal distribution as a function of $C_{ij}$ (or $B^\pm_i$) is obtained by reweighting the spin-uncorrelated Higgs boson sample with the spin weight $W_T$ defined in Sec.~\ref{sec:simulations}. Each element is scanned individually, with the remaining elements held fixed at their standard model (SM) values. A Poisson likelihood is constructed for each category, and the categories are combined by summing their contributions to $-2\ln L$ at each scan point.

Both the signal and background normalizations are fixed to their expected yields, and the background shape is fixed to the SM expectation, so that only the shape of the discriminating variable varies with the fitted parameter. Only statistical uncertainties are considered, and no nuisance parameters are included. This is an adequate approximation for a projection of this kind: the \CP measurement of Ref.~\cite{CMS:2026hvv} is strongly statistically dominated, with systematic uncertainties amounting to only around 5\% of the statistical uncertainty.

\begin{figure}[!htbp]
\includegraphics[width=\linewidth]{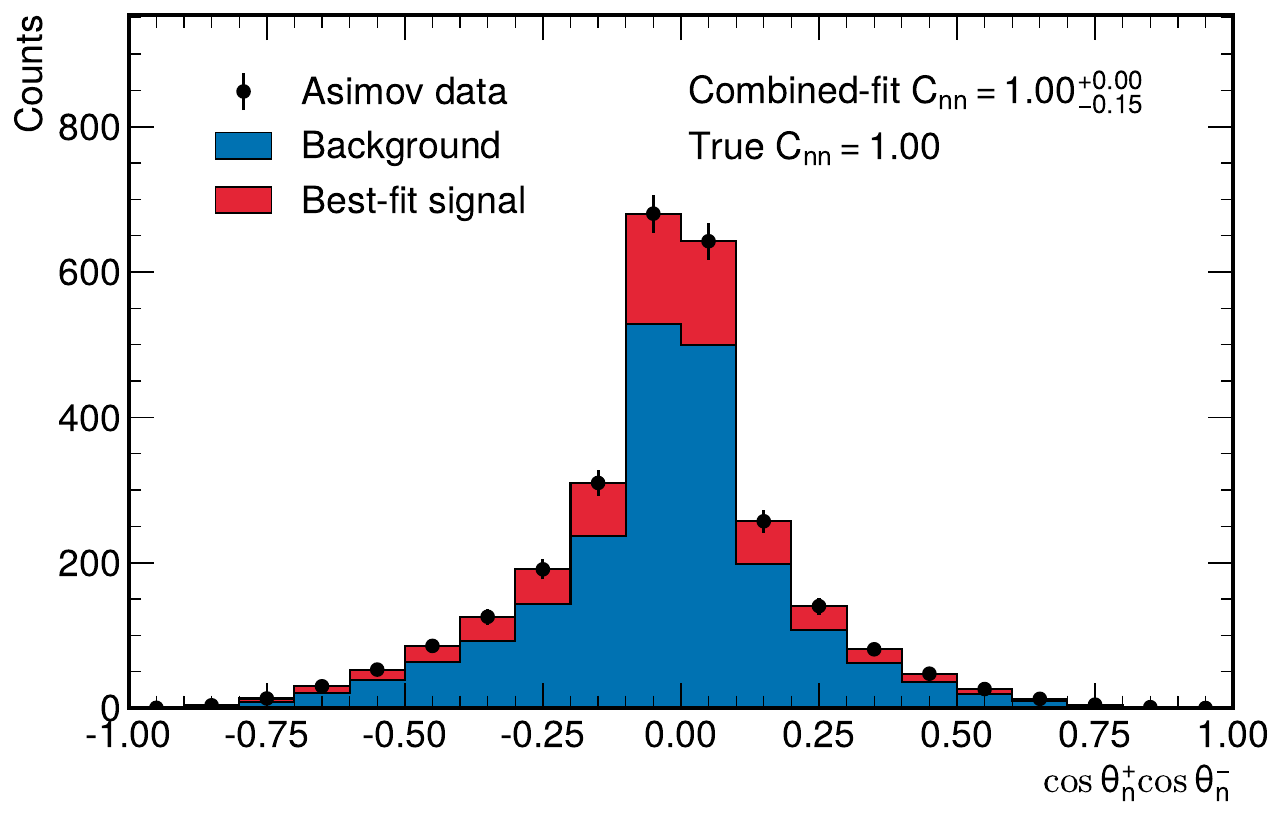} \\
\includegraphics[width=\linewidth]{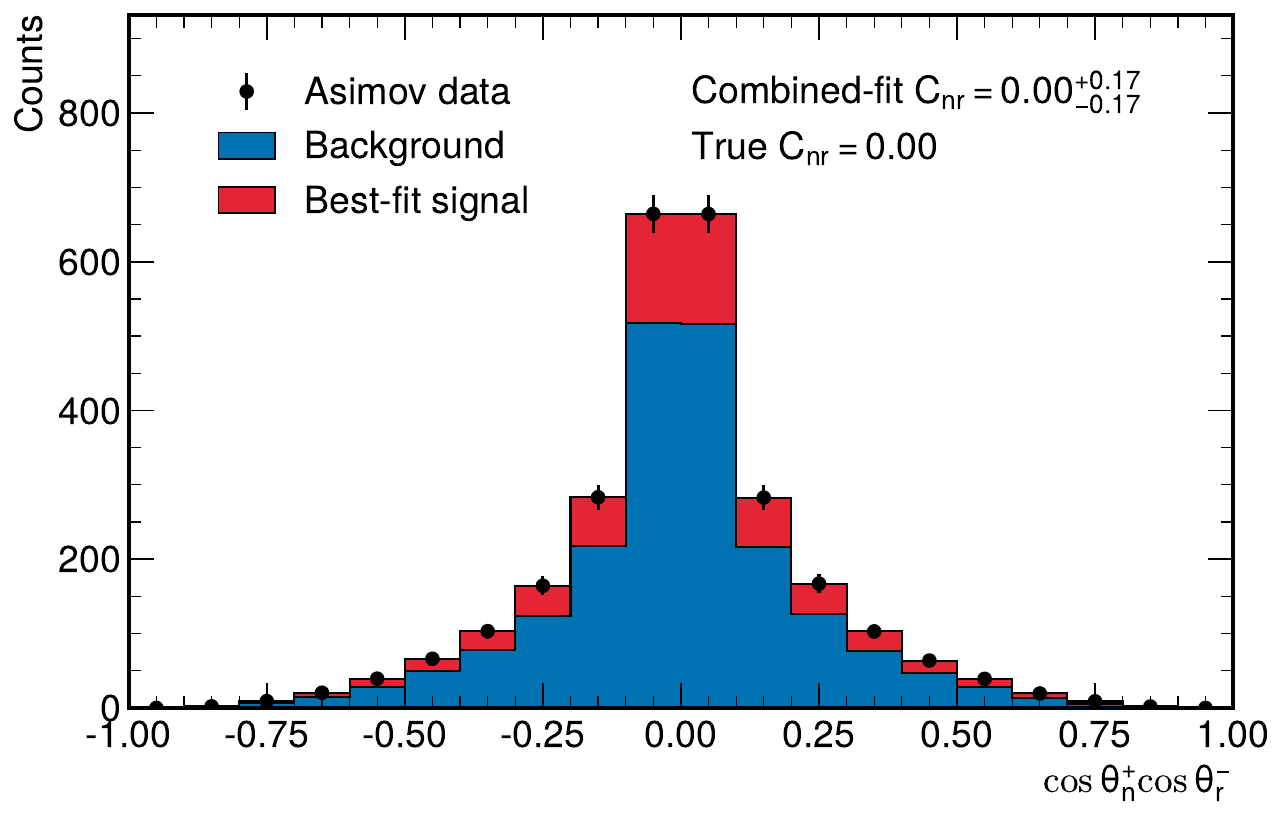}
\caption{Examples of the distributions used to extract the elements of the $\mathbf{C}$ matrix, shown for the $1\pi^\pm1\pi^0$--$1\pi^\pm1\pi^0$ category, in which both $\tau$ leptons are reconstructed in the $\tau^\mp\rightarrow\pi^\mp\pi^0\nu$ decay mode. The distributions of $\cos\theta_n^{+}\cos\theta_n^{-}$ (upper) and $\cos\theta_n^{+}\cos\theta_r^{-}$ (lower) are used to extract $C_{nn}$ and $C_{nr}$, respectively. The Asimov data are shown as black points, with the best-fit Higgs boson signal (red) stacked on the $Z/\gamma^*\rightarrow\tau\tau$ background (blue). The quoted values are those of the fit combined over all categories.}
\label{fig:fit_example}
\end{figure}

The fits are performed on an Asimov dataset constructed assuming the SM, \ie a \CP-even Higgs boson signal, together with the SM background expectation. Examples of the fitted distributions used to extract $C_{nn}$ and $C_{nr}$ are shown in Fig.~\ref{fig:fit_example}. The resulting best-fit values and uncertainties are
\begin{equation}
\begin{aligned}
\mathbf{B} &= \begin{pmatrix} 0.00\pm0.06 \\ 0.00\pm0.07 \\ 0.00\pm0.05 \end{pmatrix}, \\[4pt]
\mathbf{C} &= \begin{pmatrix}
+1.00^{+0.00}_{-0.15} & 0.00\pm0.17 & 0.00\pm0.13 \\
0.00\pm0.17 & +1.00^{+0.00}_{-0.17} & 0.00\pm0.13 \\
0.00\pm0.13 & 0.00\pm0.13 & -1.00^{+0.10}_{-0.00}
\end{pmatrix}.
\end{aligned}
\end{equation}
The $\mathbf{B}^{+}$ and $\mathbf{B}^{-}$ vectors are measured independently and are found to be equal within their uncertainties, as expected; a single averaged vector $\mathbf{B}$ is therefore quoted. The asymmetric uncertainties on the diagonal elements of $\mathbf{C}$ arise because these elements lie at the edge of their physically allowed range, $|C_{ij}|\leq1$. The uncertainties quoted are those obtained from the likelihood scan; they were cross-checked using the pseudoexperiments described below and found to agree closely.

The concurrence $\mathcal{C}$ and the quantity $m_{12}$ are then obtained from the spin density matrix reconstructed from the fitted $\mathbf{C}$ matrix and $\mathbf{B}^\pm$ vectors. For the Asimov dataset they reproduce the expected values exactly, $\mathcal{C}=1$ and $m_{12}=2$. These values are not, however, representative of what an analysis of a real, finite dataset would measure: both quantities are non-linear functions of the fitted matrix elements, and both are biased once the $C_{ij}$ carry statistical uncertainties, $\mathcal{C}$ downward and $m_{12}$ upward.

The origin of these biases follows directly from the definitions of the two quantities. From Eq.~(\ref{eqn:m12}), $m_{12}=m_1+m_2$ is the sum of the two largest eigenvalues of $\mathbf{M}=\mathbf{C}\mathbf{C}^{T}$; since $\mathrm{Tr}\,\mathbf{M}=\sum_{ij}C_{ij}^2$, this is equivalent to $m_{12}=\sum_{ij}C_{ij}^2-m_3$. In the SM the diagonal elements sit at the boundary $|C_{ii}|=1$ and can therefore only fluctuate inward, whereas the off-diagonal elements are zero and fluctuate in either direction. Non-zero off-diagonal elements spread the eigenvalue spectrum of $\mathbf{M}$, raising $m_1$ and lowering $m_3$; because $m_{12}$ retains the two largest eigenvalues and discards $m_3$, it captures the increase while remaining insensitive to the compensating decrease, and is therefore biased upward. The concurrence, in contrast, attains its maximum possible value $\mathcal{C}=1$ for the SM, so any mismeasurement of the $C_{ij}$ can only reduce it. 
As an illustration, off-diagonal fluctuations of $C_{nr} = C_{rn} = 0.17$, equal in size to the uncertainties obtained above, increase $m_{12}$ from 2.00 to 2.37 and decrease $\mathcal{C}$ from 1.00 to 0.83.

\begin{figure}[!htbp]
\includegraphics[width=\linewidth]{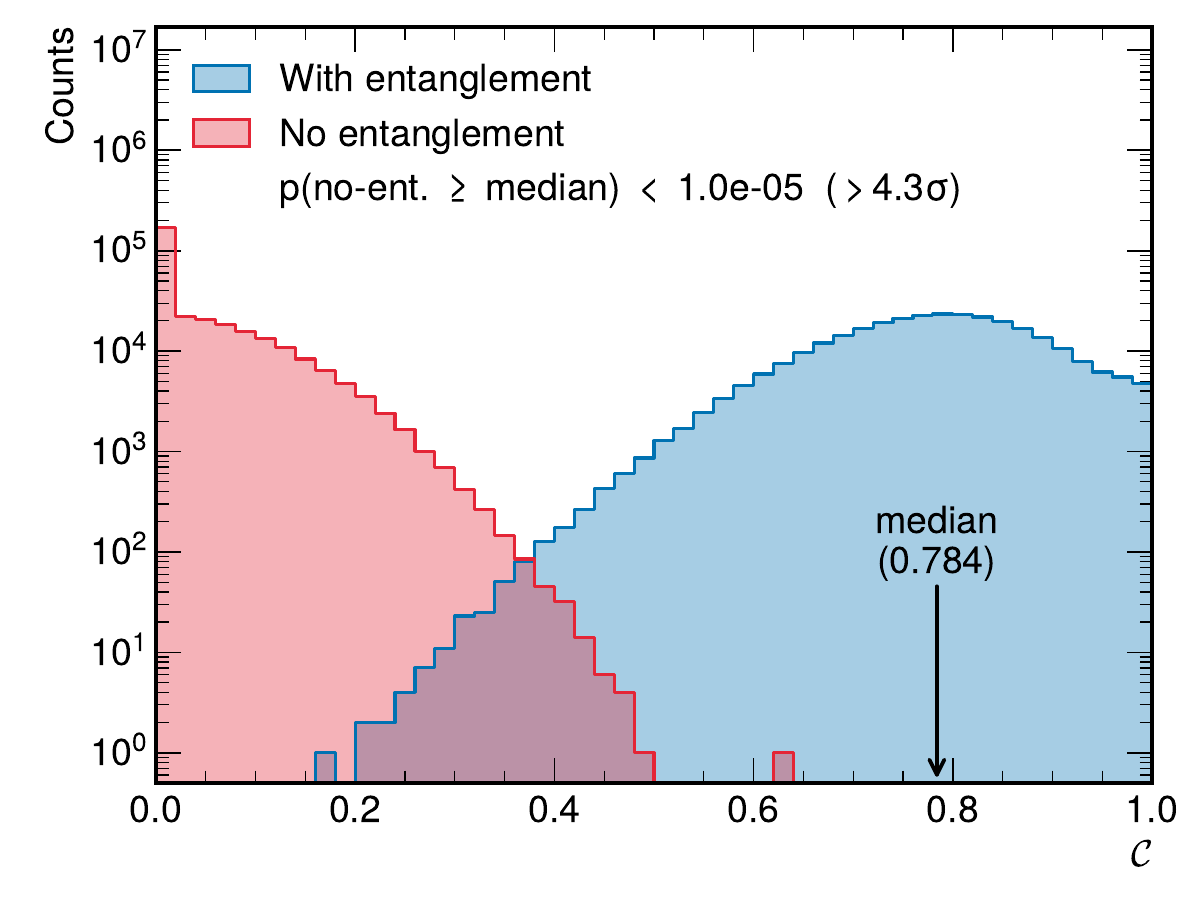} \\
\includegraphics[width=\linewidth]{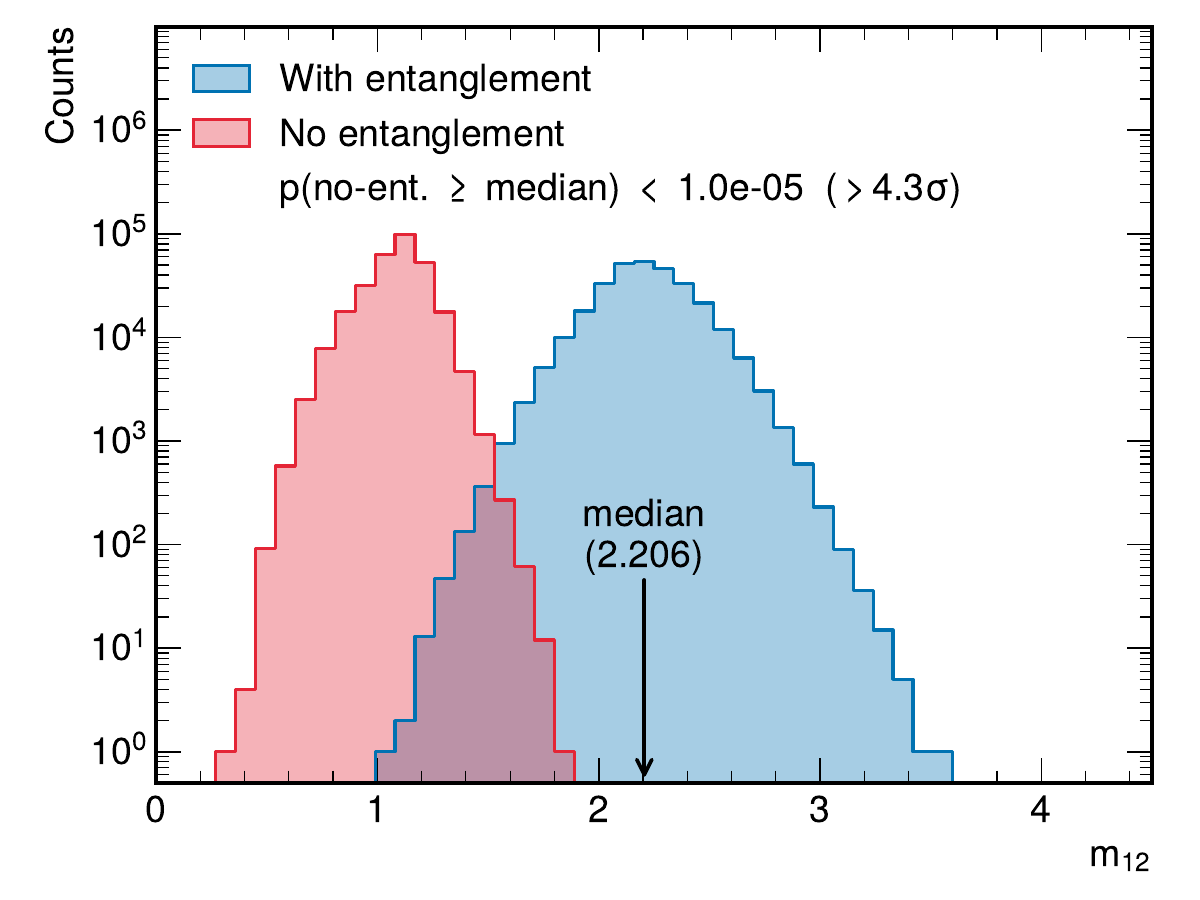}
\caption{Distributions of the measured $\mathcal{C}$ (upper) and $m_{12}$ (lower) for 300000 pseudoexperiments, generated under the with-entanglement (blue) and no-entanglement (red) hypotheses. The black arrow marks the median of the with-entanglement distribution. Taking this median as the observed value, the probability of the no-entanglement hypothesis producing a value at least as extreme, $p(\text{no-entanglement} \geq \text{median})$, is $<1.0\times10^{-5}$ at 95\% confidence level for both $\mathcal{C}$ and $m_{12}$, as no no-entanglement pseudoexperiment reaches the median, indicating a clear separation between the two hypotheses.}
\label{fig:toys}
\end{figure}

We use a set of pseudoexperiments to evaluate the expected spread of the measured values of these quantities. The signal and background yields of each category are first drawn from Poisson distributions with means set to the Asimov expectations. Events are then drawn with replacement from the simulated signal and background samples according to these yields, with signal events sampled in proportion to their spin weight $W_T$ so that the ensemble follows the hypothesis under test. Each pseudoexperiment is then fitted in exactly the same way as the Asimov dataset, and $\mathcal{C}$ and $m_{12}$ are computed from the resulting $\mathbf{C}$ matrix and $\mathbf{B}^\pm$ vectors. The distributions of these measured values are shown for 300000 pseudoexperiments in Fig.~\ref{fig:toys} (blue histograms).

We also compare them with the distributions obtained under a null hypothesis in which the $\tau$ spins are classically correlated but not entangled, constructed by retaining the longitudinal spin correlation of the SM while setting the transverse correlations to zero:
\begin{equation}
    C_{ij} =
\begin{cases}
-1 & i=j=k \\
0 & \text{otherwise}
\end{cases}.
\end{equation}
The $\tau$ spins therefore remain anti-aligned along $\mathbf{k}$, and the two hypotheses are indistinguishable if both spins are analyzed along that axis; they differ only in the transverse correlations, in which the quantum coherence of the SM state resides. The resulting state is separable, with $\mathcal{C}=0$, and gives $m_{12}=1$, exactly at the threshold below which no violation of Bell's inequality is implied. Its pseudoexperiment distributions are shown as the red histograms in Fig.~\ref{fig:toys}; since the concurrence is clipped at zero by the $\max(0,\ldots)$ in Eq.~(\ref{eqn:concurrence}), this ensemble accumulates at $\mathcal{C}=0$.

Finally, we assess the expected $p$-value for separating the two hypotheses. The observed values of $\mathcal{C}$ and $m_{12}$ are taken to be the medians of the with-entanglement ensemble, and the $p$-value is estimated as the fraction of no-entanglement pseudoexperiments yielding a value at least as large. None of the 300000 no-entanglement pseudoexperiments reaches the with-entanglement median for either observable, so we obtain $p<1.0\times10^{-5}$ at 95\% confidence level for both $\mathcal{C}$ and $m_{12}$, corresponding to one-sided significances of more than $4.3\sigma$. These bounds are set by the number of pseudoexperiments generated, and the true separation is expected to be larger. We therefore conclude that the HL-LHC should be able to distinguish between the two hypotheses, barring an unfavorable statistical fluctuation.

\section{Application to \CP Measurements}
\label{sec:cp}
The \CP nature of the Yukawa coupling between the Higgs boson and $\tau$ leptons affects the spin correlations between the two $\tau$ leptons produced in $H\to\tau\tau$ decays. These correlations modify the angular properties of the $\tau$ decay, allowing the determination of the \CP mixing angle $\alpha$ from the acoplanarity \phicp, defined as the angle between $\tau$ lepton decay planes in the Higgs boson rest frame. The \phicp distributions exhibit a phase shift which depends on the value of $\alpha$, with the \CP-even and \CP-odd scenarios having opposite modulation.
The effective \CP mixing angle $\alpha$ has been measured by the CMS and ATLAS collaborations~\cite{CMS:2021sdq,ATLAS:2022akr,CMS:2026hvv}. 
However, the measurement is challenging as the full $\tau$ lepton momenta are unknown due to the presence of neutrinos in every final state, rendering the accurate reconstruction of \phicp difficult. 
 Improvements to the approximate methods which were previously employed could reduce the uncertainty on the measurement, which is required to precisely probe new physics scenarios.

Three methods have been used previously to determine \phicp: (i) using the impact parameter vector of the charged decay product in place of the unmeasured $\tau$ lepton direction; (ii) determining the decay plane of the intermediate $\rho$ resonance in decays involving $\pi^0$s; and (iii) using the polarimetric vector.
The polarimetric vector method provides optimal sensitivity to $\alpha$, however this could previously only be applied to $\tau$ lepton decays involving three charged particles, where the secondary decay vertex provides a constraint on the direction of the $\tau$ momentum. This technique relies heavily on the secondary vertex reconstruction, while \texttt{TauPolaris} can exploit the full reconstructed event information, improving the polarimetric vector determination in these modes as well.

The neutrino estimation presented in this work extends the usage of the polarimetric vector method to every $\tau$ lepton decay topology, determining \phicp from the angle between planes spanned by each $\tau$ lepton and its polarimetric vector. The full $\tau$ lepton momentum is determined by combining the regressed neutrino with the visible decay products and the polarimetric vector is obtained as described in Sec.~\ref{sec:models}. The correct modeling of correlations between the different components of the $\tau$ momentum, and the accurate determination of the direction, are crucial to achieving good \phicp reconstruction, and therefore increasing sensitivity to $\alpha$.

Similarly to the spin variables in Sec.~\ref{sec:entanglement}, the \phicp values are estimated using 200 samples from the learned $p\!\left(\bm{\nu} \mid \mathbf{c}\right)$, except that the circular mean~\cite{Mardia:1999directional} is used to account for the periodicity of \phicp.

The performance of the \texttt{TauPolaris}, transformer, and the approximate methods as described in Ref.~\cite{CMS:2026hvv} are quantified by the asymmetry observed in \phicp distributions. This asymmetry is defined as:
\begin{equation}
    A = \sqrt{\sum_{i=1}^{N_{bins}} (N_i^{\textrm{Even}} - N_i^{\textrm{Odd}})^2 }
\end{equation}
where
$N_i^{\textrm{Even}}$ ($N_i^{\textrm{Odd}}$) are the number of \CP-even (CP-odd) events in each bin, with both distributions normalized to unity. This measure is used because it is proportional to the sensitivity to $\alpha$ that a binned likelihood fit to the \phicp distribution would achieve: the background is uniformly distributed in \phicp and exceeds the signal in all the categories considered, so the statistical uncertainty is approximately equal in every bin. The ratio of $A$ between two reconstruction methods therefore gives the relative improvement in sensitivity to $\alpha$.

\begin{figure*}[!htbp]
\includegraphics[width=0.32\linewidth]{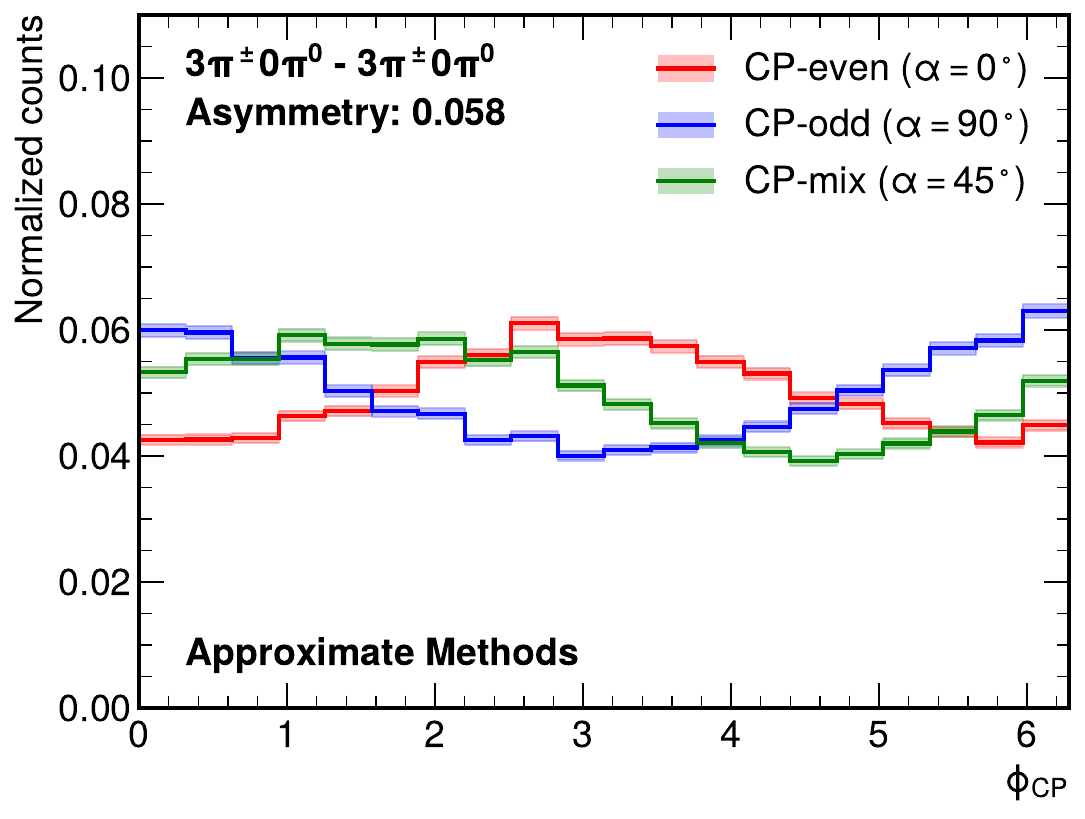}
\includegraphics[width=0.32\linewidth]{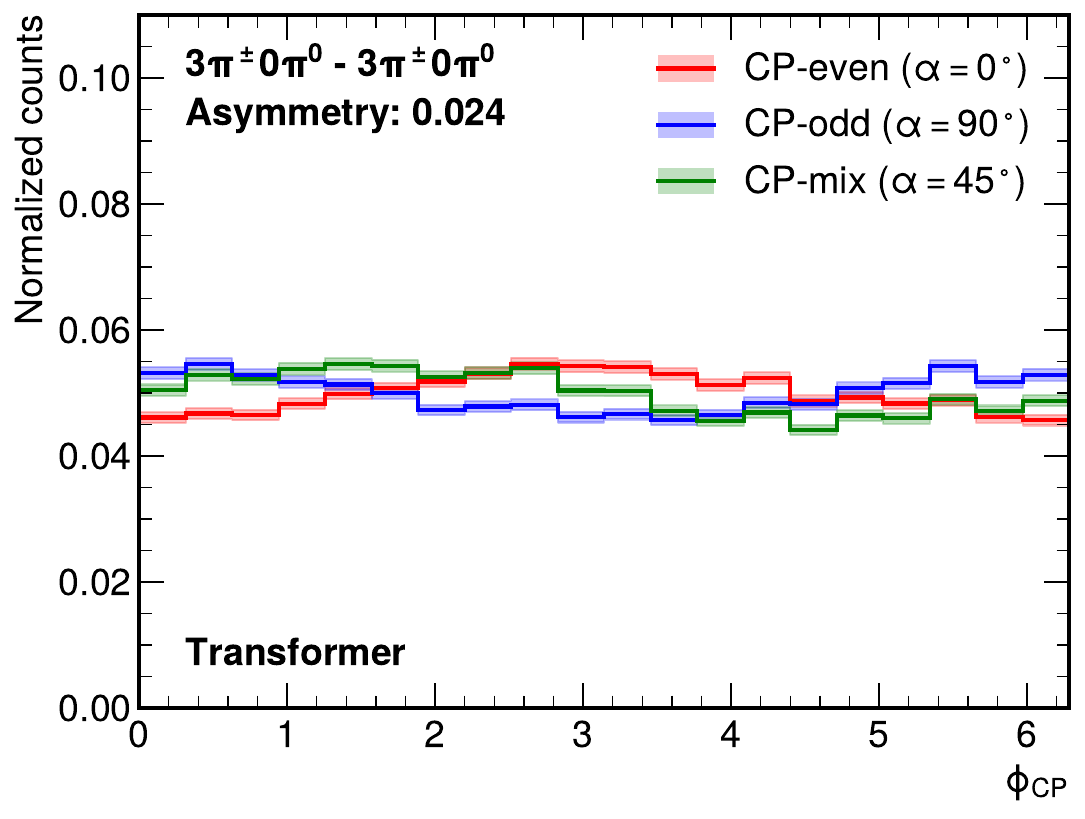}
\includegraphics[width=0.32\linewidth]{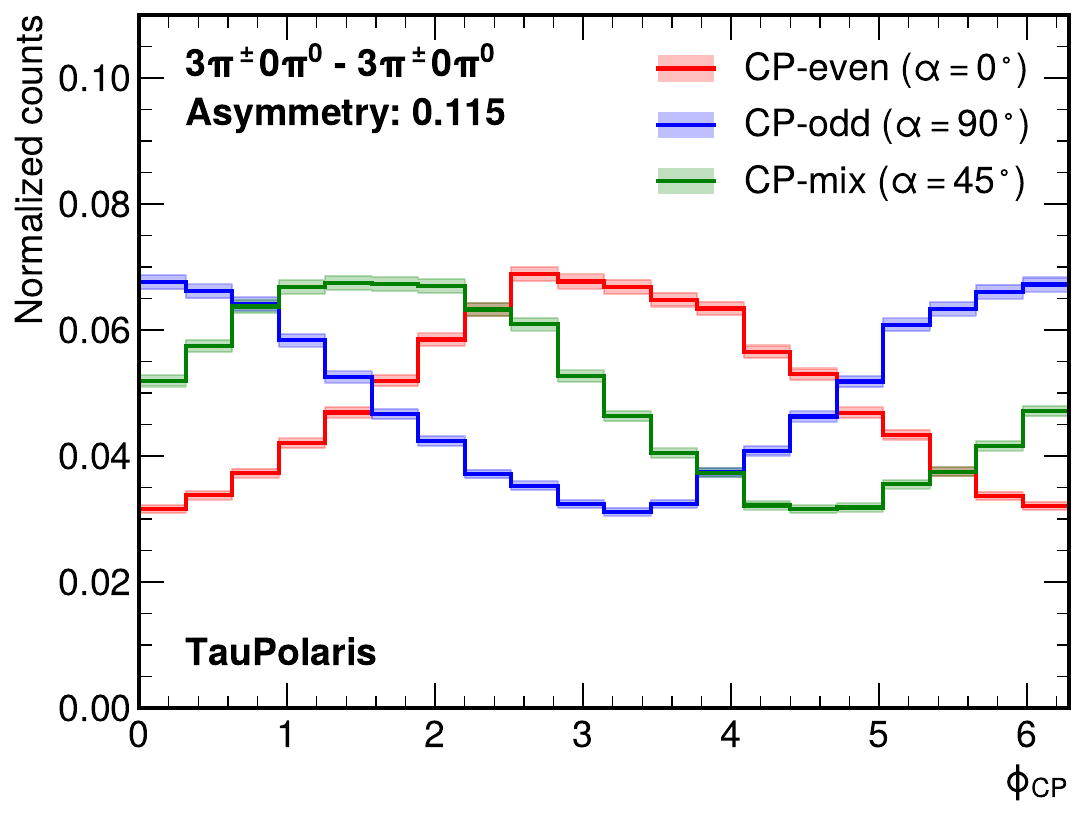}
\caption{The distribution of the \phicp observable in the $3\pi^\pm0\pi^0-3\pi^\pm0\pi^0$ final state for the approximate method (left), stand-alone transformer (middle) and \texttt{TauPolaris} (right) for different \CP scenarios. The \texttt{TauPolaris} algorithm achieves a 97\% larger asymmetry between \CP-even and \CP-odd scenarios than the approximate method, and 380\% larger asymmetry than the stand-alone transformer.}
\label{fig:phiCPDM10DM10}
\end{figure*}

Notable improvements are achieved in all decay modes.
The \texttt{TauPolaris} algorithm has significantly increased the \CP separation in decay modes involving three charged hadrons, as shown in Fig.~\ref{fig:phiCPDM10DM10}, with improvements of 97\% when both $\tau$ leptons decay to $3\pi^\pm0\pi^0$.
These improvements cannot be achieved with the stand-alone transformer approach, due to less accurate modeling of the neutrino direction, which is crucial for correctly determining the polarimetric vector. The previous approximate techniques perform better than the transformer in the majority of decay topologies, as they rely on visible $\tau$ information instead of poorly reconstructed neutrinos.

\begin{figure*}[!htbp]
\includegraphics[width=0.32\linewidth]{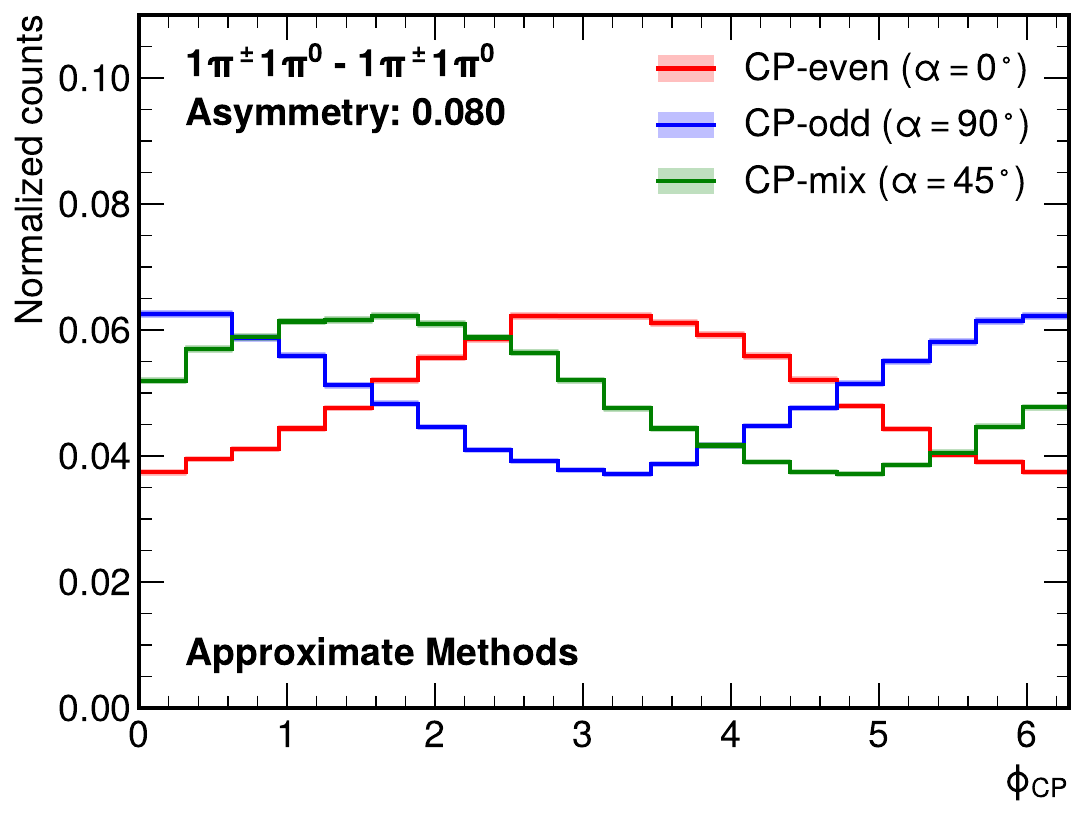}
\includegraphics[width=0.32\linewidth]{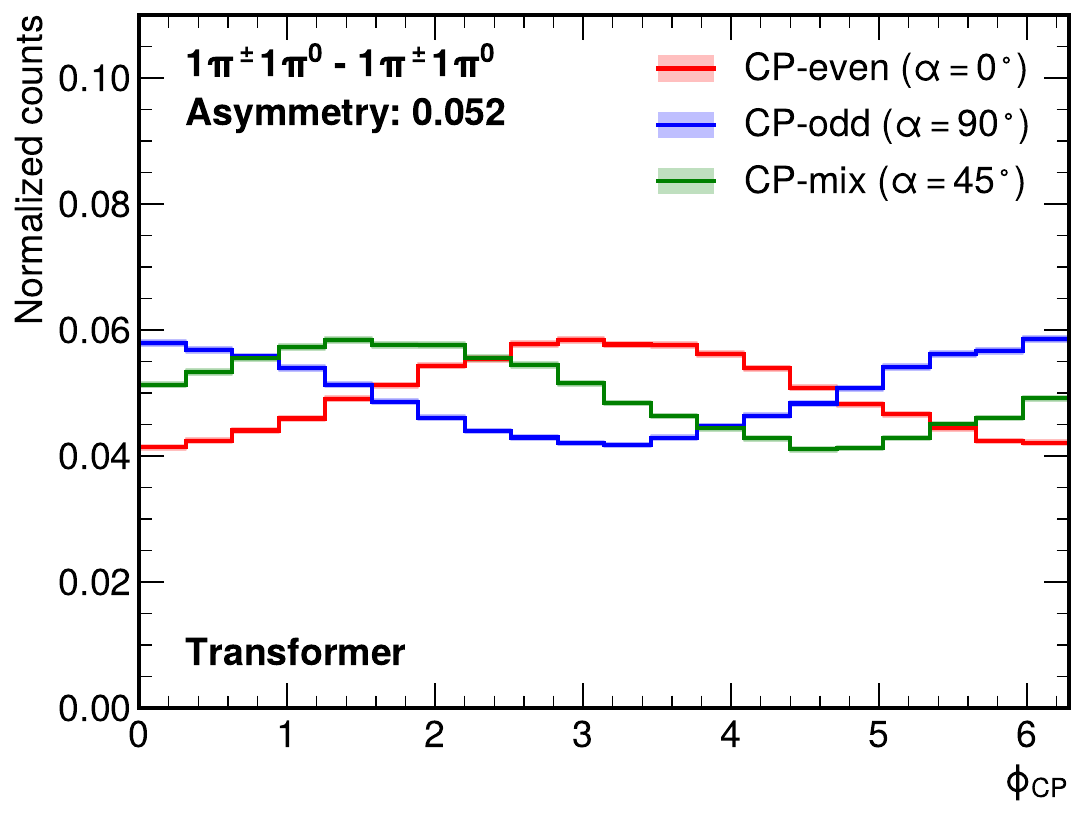}
\includegraphics[width=0.32\linewidth]{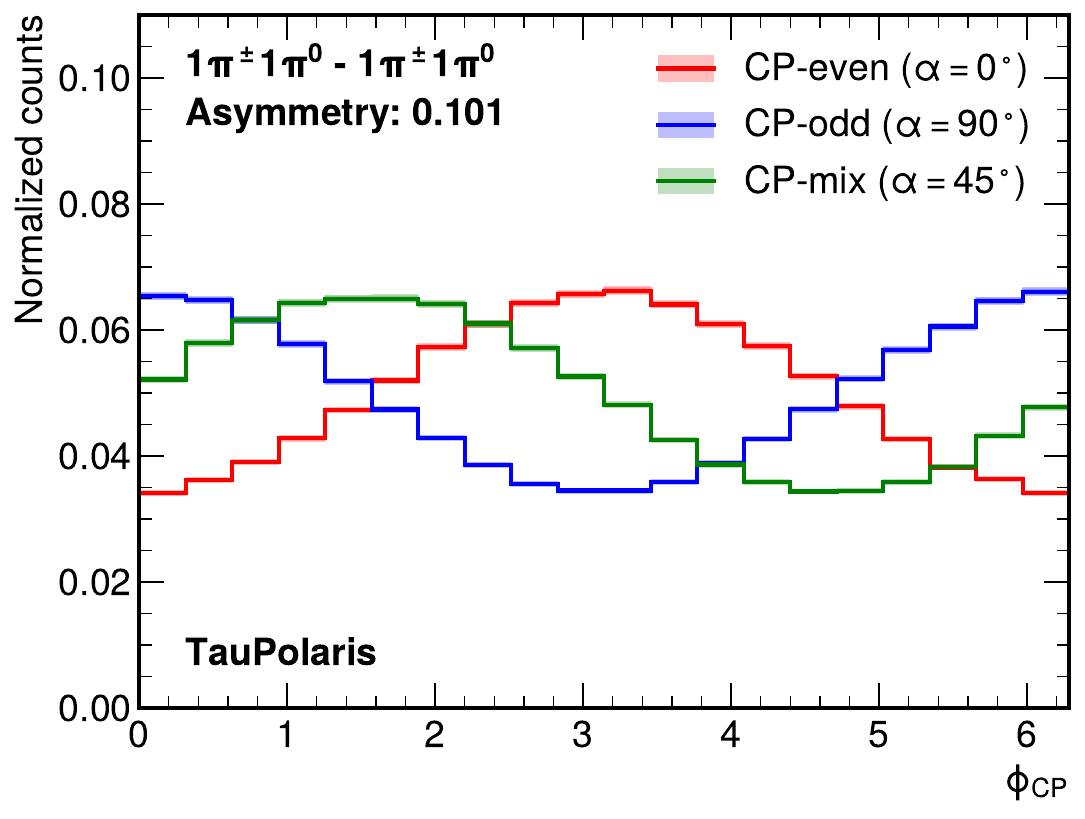}
\caption{The distribution of the \phicp observable in the $1\pi^\pm1\pi^0$--$1\pi^\pm1\pi^0$ channel (which was the most sensitive final state in the previous CMS measurement~\cite{CMS:2026hvv}) for the approximate method (left), stand-alone transformer (middle) and \texttt{TauPolaris} (right) for different \CP scenarios. The \texttt{TauPolaris} algorithm achieves a 27\% larger asymmetry between \CP-even and \CP-odd scenarios than the approximate methods, and 94\% larger asymmetry than the stand-alone transformer.}
\label{fig:phiCPDM1DM1}
\end{figure*}

Improvements are also made in the final states which were found to contribute most to the $\alpha$ sensitivity in the previous CMS measurement~\cite{CMS:2026hvv}, as shown in Fig.~\ref{fig:phiCPDM1DM1} for the $1\pi^\pm1\pi^0$--$1\pi^\pm1\pi^0$ channel, where the asymmetry is increased by 27\% relative to the previous approximate method.

Furthermore, the estimation of the neutrino produced in $3\pi^\pm1\pi^0$ decays allows the experimental reconstruction of the \phicp angle for decays involving this topology for the first time. The reconstructed neutrino is combined with the visible decay products to determine the polarimetric vector, following the formalism implemented in \texttt{TAUOLA}~\cite{Jadach:1990mz,Jezabek:1991qp,Jadach:1993hs,Davidson:2010rw}.

A comparison of the asymmetry achieved for all the decay modes is given in Table~\ref{tab:cp_asymmetry}. 
The percentage improvement in asymmetry of the \texttt{TauPolaris} algorithm compared to the approximate ($\Delta_\textrm{Approx.}$) and transformer methods ($\Delta_\textrm{Transf.}$) is also included.

In order to quantify the overall performance improvement, we define a summary statistic:
\begin{equation}
    A_{\text{total}} = \sqrt{\frac{\sum_{c}  N_c \cdot A_c^2}{\sum_c N_c}}
\end{equation}
where $A_c$ and $N_c$ are the asymmetry and event count of decay mode combination $c$. This quantity accounts for the sensitivity scaling with the square root of the number of events in each final state.
The \texttt{TauPolaris} algorithm achieves $A_{\text{total}}=0.0707$, corresponding to an improvement of 36\% compared with the approximate methods (0.0521), and 85\% compared with the stand-alone transformer (0.0383). 
Reconstruction of \phicp was also attempted using $\tau$ lepton momenta estimated by \texttt{SVFit}. However, due to poor estimation of the $\tau$ lepton direction, this approach was not competitive, achieving  $A_{\text{total}}=0.0116$, and is therefore not shown.

\begin{table*}[!htbp]
  \centering
  \setlength{\tabcolsep}{7pt}
  \caption{Asymmetry of the \phicp distributions in each reconstructed decay-mode category of the \tauhtauh and \taultauh channels, for the \texttt{TauPolaris} estimate, the stand-alone transformer regressor, and the previous approximate reconstruction methods. A larger asymmetry corresponds to greater separation between the \CP hypotheses.  $\Delta_\text{Approx.}$  represents the percentage improvement achieved with \texttt{TauPolaris} compared with the approximate method, while $\Delta_\text{Transf.}$ represents the \texttt{TauPolaris} improvement compared with the stand-alone transformer.}
  \label{tab:cp_asymmetry}
\begin{tabular}{l c c c c c}
    \toprule
    Decay modes & Approximate & Transformer & \texttt{TauPolaris} & $\Delta_\text{Approx.}$ (\%) & $\Delta_\text{Transf.}$ (\%) \\
    \midrule
    $1\pi^\pm0\pi^0$--$1\pi^\pm0\pi^0$ & 0.084 & 0.090 & 0.090 & $+7$   & $+0$   \\
    $1\pi^\pm0\pi^0$--$1\pi^\pm1\pi^0$ & 0.077 & 0.061 & 0.088 & $+16$  & $+45$  \\
    $1\pi^\pm0\pi^0$--$1\pi^\pm2\pi^0$ & 0.054 & 0.040 & 0.065 & $+20$  & $+63$  \\
    $1\pi^\pm0\pi^0$--$3\pi^\pm0\pi^0$ & 0.062 & 0.039 & 0.096 & $+53$  & $+147$ \\
    $1\pi^\pm1\pi^0$--$1\pi^\pm1\pi^0$ & 0.080 & 0.052 & 0.101 & $+27$  & $+94$  \\
    $1\pi^\pm1\pi^0$--$1\pi^\pm2\pi^0$ & 0.056 & 0.034 & 0.074 & $+32$  & $+118$ \\
    $1\pi^\pm1\pi^0$--$3\pi^\pm0\pi^0$ & 0.066 & 0.031 & 0.108 & $+64$  & $+248$ \\
    $1\pi^\pm2\pi^0$--$1\pi^\pm2\pi^0$ & 0.040 & 0.022 & 0.056 & $+39$  & $+155$ \\
    $1\pi^\pm2\pi^0$--$3\pi^\pm0\pi^0$ & 0.047 & 0.019 & 0.080 & $+72$  & $+323$ \\
    $3\pi^\pm0\pi^0$--$3\pi^\pm0\pi^0$ & 0.058 & 0.024 & 0.115 & $+97$  & $+380$ \\
    $1\pi^\pm0\pi^0$--$3\pi^\pm1\pi^0$ & --    & 0.019 & 0.054 & --     & $+182$ \\
    $1\pi^\pm1\pi^0$--$3\pi^\pm1\pi^0$ & --    & 0.015 & 0.062 & --     & $+311$ \\
    $1\pi^\pm2\pi^0$--$3\pi^\pm1\pi^0$ & --    & 0.011 & 0.045 & --     & $+311$ \\
    $3\pi^\pm0\pi^0$--$3\pi^\pm1\pi^0$ & --    & 0.011 & 0.067 & --     & $+514$ \\
    $3\pi^\pm1\pi^0$--$3\pi^\pm1\pi^0$ & --    & 0.009 & 0.042 & --     & $+364$ \\
    \midrule
    \taul--$1\pi^\pm0\pi^0$ & 0.041 & 0.044 & 0.044 & $+7$   & $+1$   \\
    \taul--$1\pi^\pm1\pi^0$ & 0.044 & 0.036 & 0.052 & $+16$  & $+44$  \\
    \taul--$1\pi^\pm2\pi^0$ & 0.032 & 0.022 & 0.039 & $+23$  & $+77$  \\
    \taul--$3\pi^\pm0\pi^0$ & 0.036 & 0.024 & 0.057 & $+57$  & $+137$ \\
    \taul--$3\pi^\pm1\pi^0$ & --    & 0.003 & 0.032 & --     & $+963$ \\
    \bottomrule
\end{tabular}
\end{table*}

\begin{figure*}[!htbp]
\includegraphics[width=0.32\linewidth]{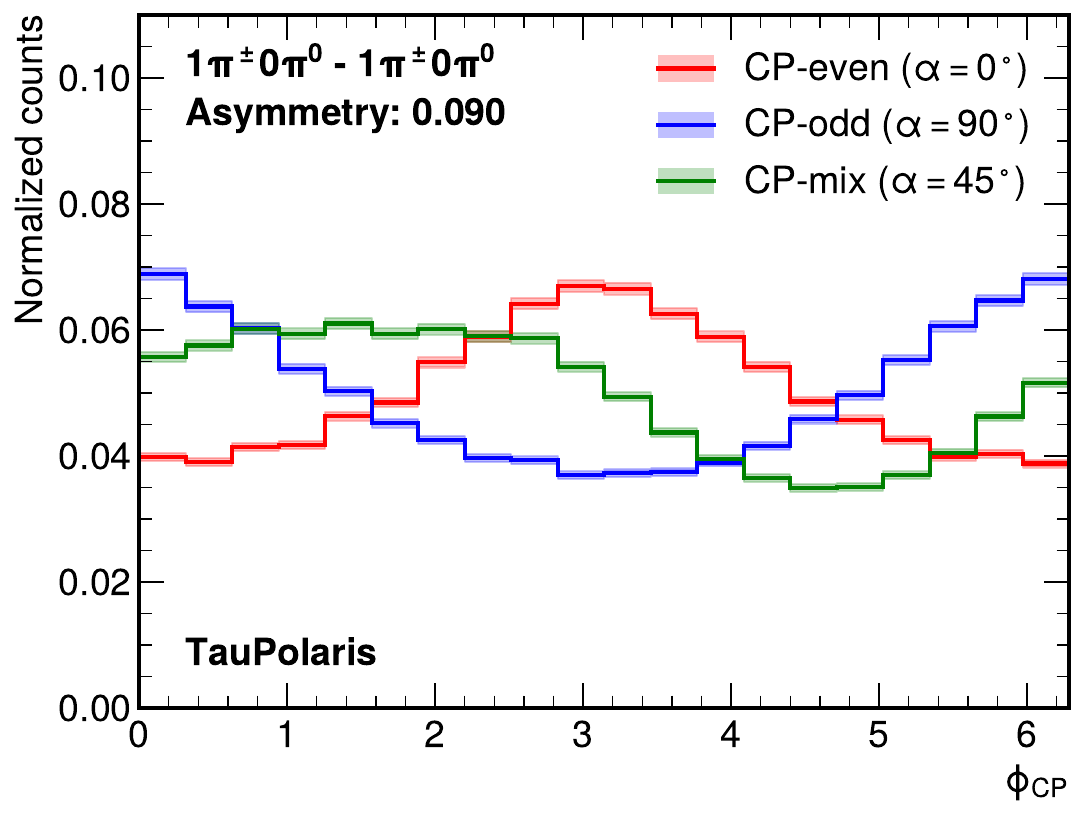}
\includegraphics[width=0.32\linewidth]{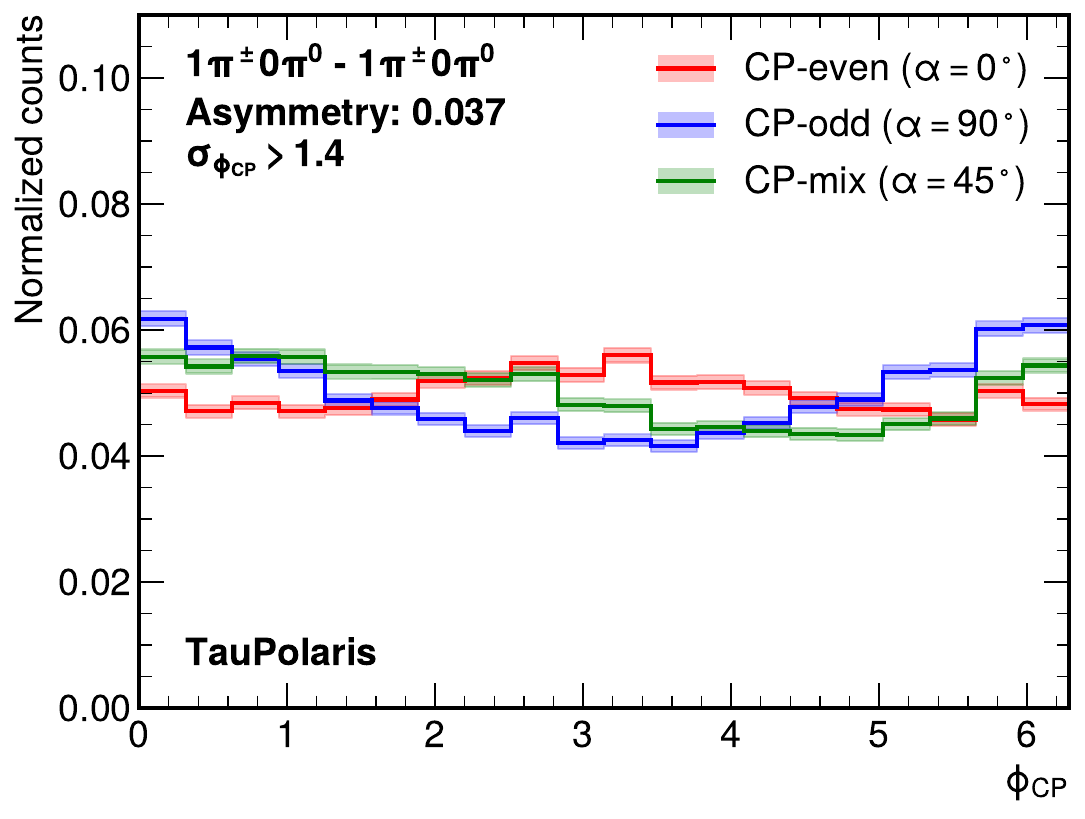}
\includegraphics[width=0.32\linewidth]{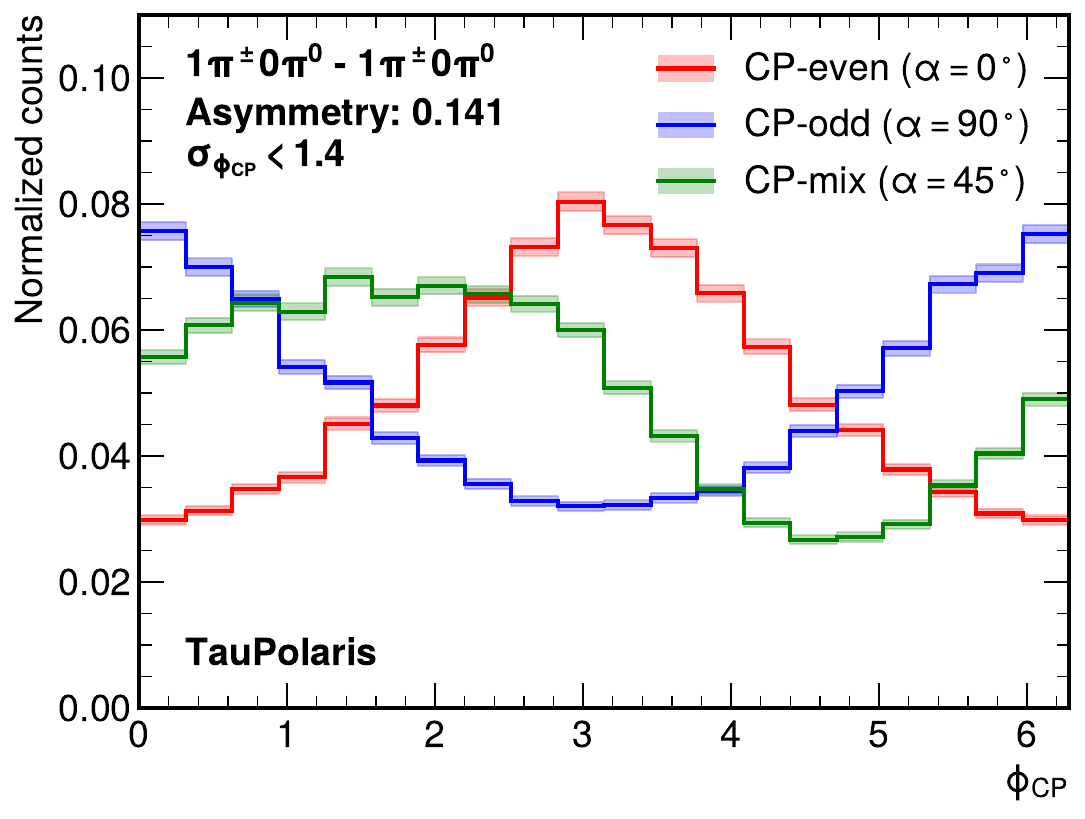}
\caption{\phicp regression performance for the $1\pi^\pm0\pi^0$--$1\pi^\pm0\pi^0$ channel for events without any selection on $\sigma_{\phicp}$ (left), events with $\sigma_{\phicp}>1.4$ (center), and events with $\sigma_{\phicp}<1.4$ (right).}
\label{fig:phicp_uncert}
\end{figure*}

As \texttt{TauPolaris} learns the full probability density, it can also be used to estimate an uncertainty on \phicp due to the estimation of the neutrinos, $\sigma_{\phicp}$. This is done by computing the circular standard deviation~\cite{Mardia:1999directional} of the 200 \phicp samples per event that are used to obtain the central \phicp value.
This can then be used to select events where \phicp is well reconstructed, which can improve the sensitivity of a \CP measurement. An example is shown for the $1\pi^\pm0\pi^0$--$1\pi^\pm0\pi^0$ channel in Fig.~\ref{fig:phicp_uncert}, where the \phicp distribution before any cut (left) is compared to the distributions after requiring $\sigma_{\phicp}>1.4$ (center) and $\sigma_{\phicp}<1.4$ (right). The value of the cut is chosen to split the events into two samples with roughly equal event populations. The asymmetry is significantly better for the sample with smaller $\sigma_{\phicp}$ compared to the sample with larger uncertainties (0.141 and 0.037, respectively). We note that an optimized event selection or categorization based on $\sigma_{\phicp}$ could increase the sensitivity to $\alpha$ beyond the 36\% improvement reported above. However, this would require a full optimization that also accounts for the impact on the background acceptance, which we leave to future work.

\section{Application to Background Suppression}
\label{sec:spin}
We can also use the reconstructed polarimetric vectors as a tool for suppressing backgrounds.
As an example, we consider an analysis targeting $H\rightarrow\tau\tau$ decays where the largest background is $Z\rightarrow\tau\tau$.
The expected $\mathbf{B}^\pm$ and $\mathbf{C}$ for the on-shell $Z\rightarrow\tau\tau$ process are~\cite{Zhang:2025mmm}:
\begin{equation}
\mathbf{C} \approx 
\begin{pmatrix}
0.5 & 0 & 0 \\
0 & -0.5 & 0 \\
0 & 0 & 1
\end{pmatrix},~\text{and}~
\mathbf{B}^\pm \approx 
\begin{pmatrix}
0  \\
0 \\
0.15 
\end{pmatrix}.
\end{equation}

A comparison with the $\mathbf{C}$ matrix expected for a SM Higgs boson (Eq.~(\ref{eqn:C_higgs}) with $\alpha=0^\circ$) suggests using $\cos\theta_n^{+}\cos\theta_n^{-}$, $\cos\theta_r^{+}\cos\theta_r^{-}$, and $\cos\theta_k^{+}\cos\theta_k^{-}$ to differentiate between signal and background through the differences in the spin correlations, and the $\cos\theta_k^{\pm}$ variables to target the differences in the $B_{k}^\pm$ elements.
However, we do not want to assume a particular value of $\alpha$, which is itself a quantity the LHC will measure. Instead of $\cos\theta_n^{+}\cos\theta_n^{-}$ and $\cos\theta_r^{+}\cos\theta_r^{-}$ we therefore define the variable $\cos\theta_n^{+}\cos\theta_n^{-}-\cos\theta_r^{+}\cos\theta_r^{-}$, which exploits the fact that $C_{nn}$ and $C_{rr}$ are always equal for Higgs boson events, whereas for $Z$ boson events they have opposite signs.
For each event, the spin variables are estimated as the mean over the flow samples, as described in Sec.~\ref{sec:entanglement}.

\begin{figure*}[!htbp]
\includegraphics[width=0.49\linewidth]{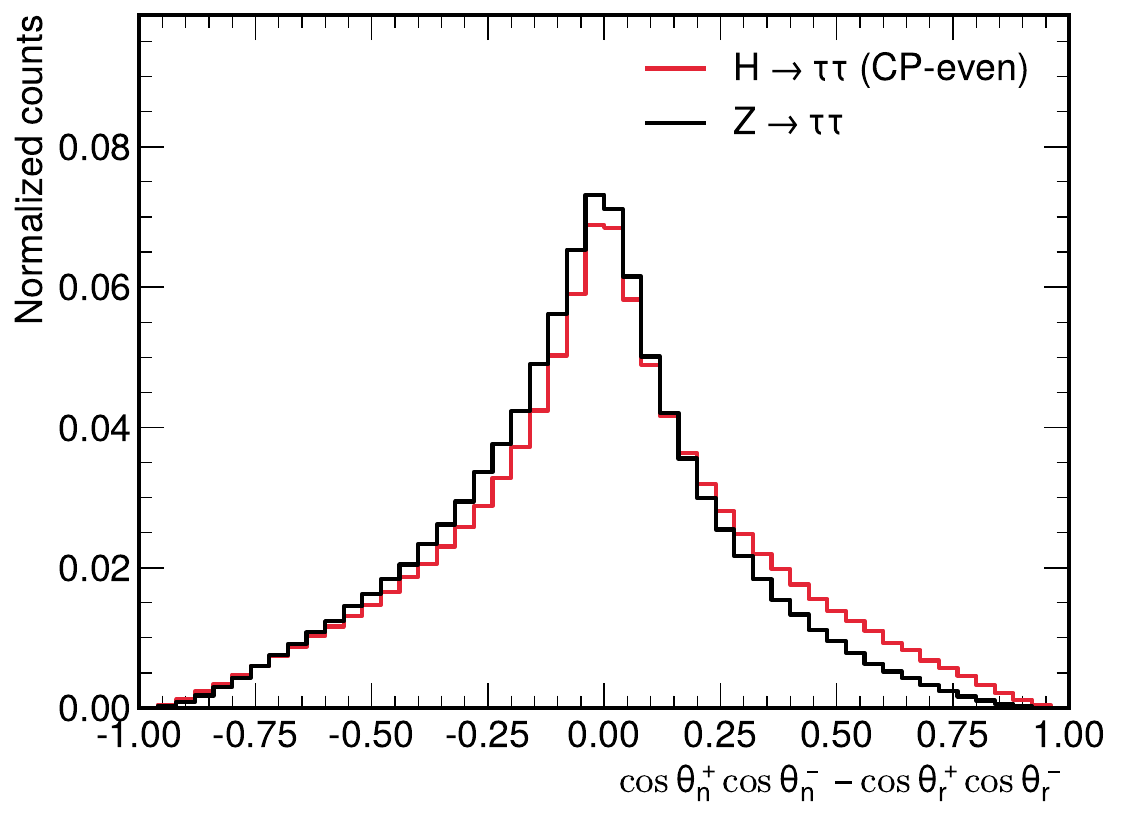}
\includegraphics[width=0.49\linewidth]{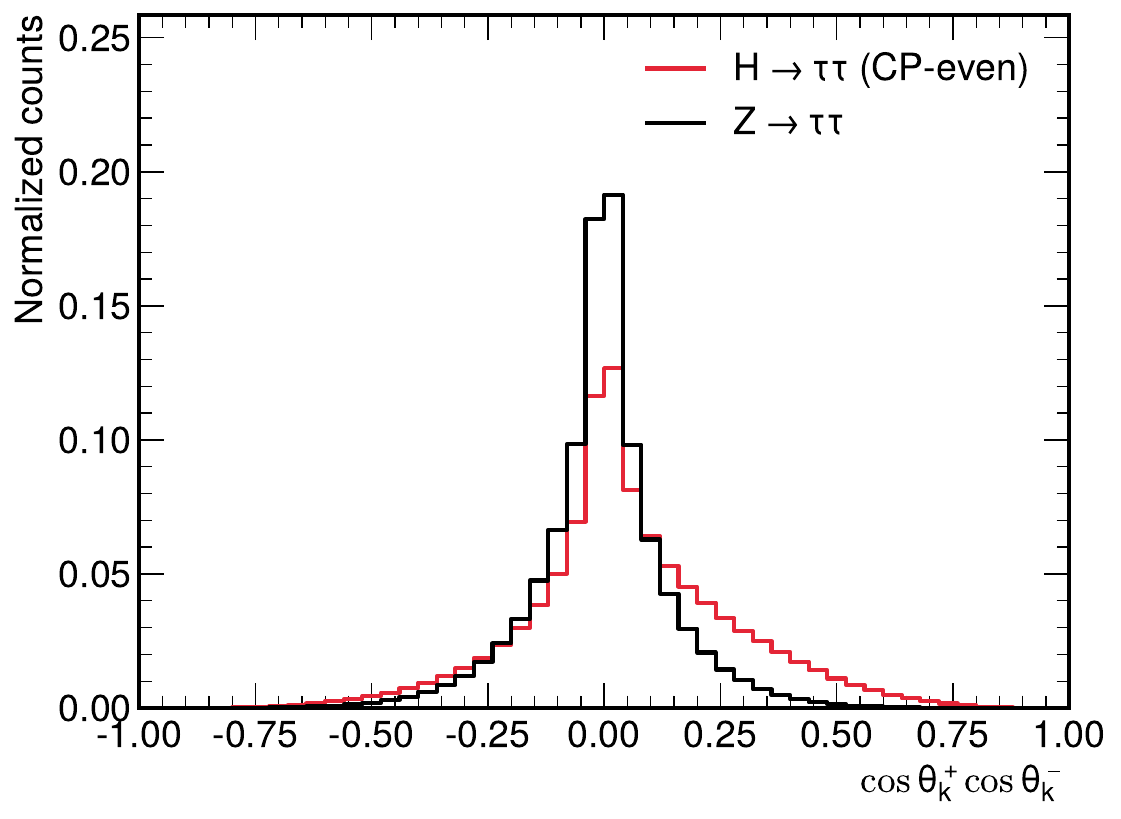} \\
\includegraphics[width=0.49\linewidth]{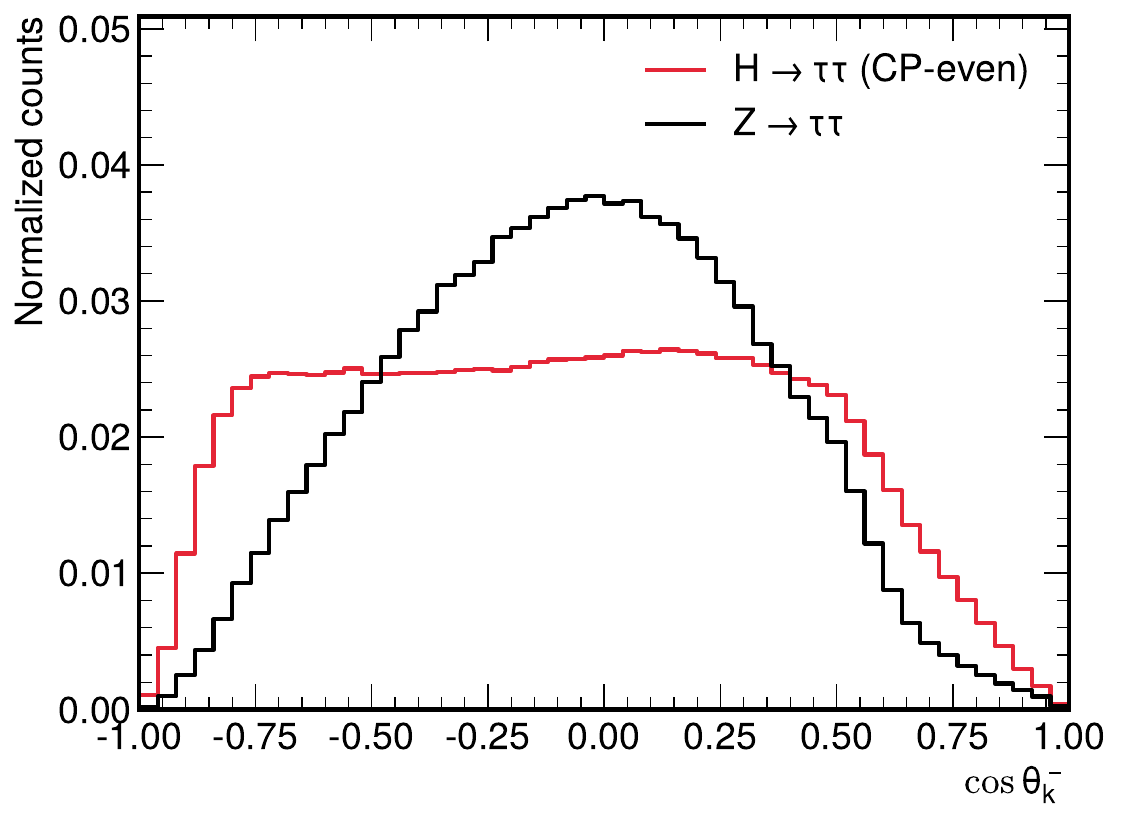}
\includegraphics[width=0.49\linewidth]{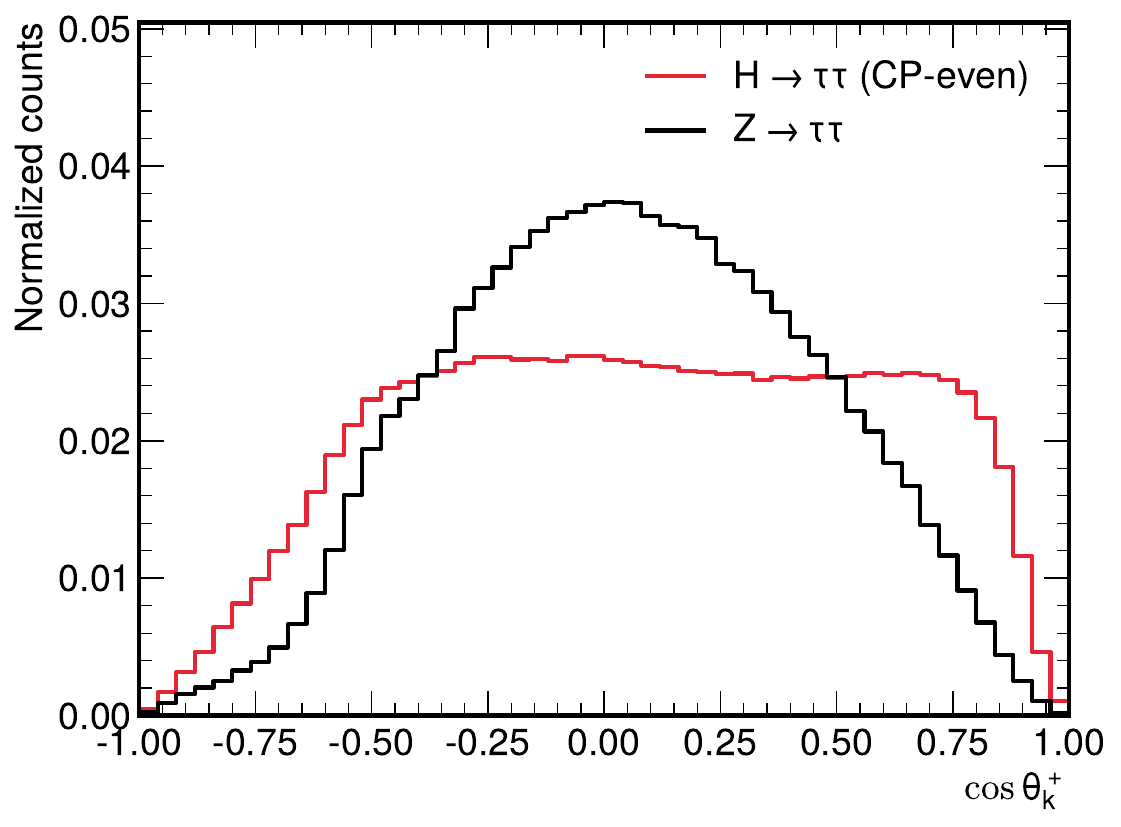}
\caption{Distributions of variables defined from the reconstructed $\tau$ lepton polarimetric vectors which can be used to differentiate between Higgs boson events (red) and $Z$ boson events (black). Shown are $\cos\theta_n^{+}\cos\theta_n^{-}-\cos\theta_r^{+}\cos\theta_r^{-}$ (upper left), $\cos\theta_k^{+}\cos\theta_k^{-}$ (upper right), $\cos\theta_k^{-}$ (lower left), and $\cos\theta_k^{+}$ (lower right). The distributions are normalized to unity.}
\label{fig:spin}
\end{figure*}

Figure~\ref{fig:spin} shows the distributions of each of the four variables, comparing Higgs boson events with $Z$ boson events in the fully hadronic \tauhtauh channel. The $1\pi^\pm0\pi^0$, $1\pi^\pm1\pi^0$, and $3\pi^\pm0\pi^0$ reconstructed decay modes are included.
Only the \CP-even Higgs boson is displayed, but the distributions are unchanged for the \CP-odd and mixed-CP scenarios: the $\alpha$ dependence of $\mathbf{C}$ amounts to a rotation in the $n$--$r$ plane, and all four variables are constructed to be invariant under it.
A reasonable separation between the two processes is observed, with $\cos\theta_k^{+}\cos\theta_k^{-}$ providing the strongest discrimination, as expected given that $C_{kk}$ has opposite signs for the two processes. These variables could therefore be used as additional handles for reducing backgrounds in Higgs boson searches at the LHC.

\section{Summary}
\label{sec:summary}
We have presented \texttt{TauPolaris}, a tool for reconstructing the polarimetric vectors of $\tau$
leptons using conditional normalizing flows.
The momenta of the undetected neutrinos are estimated conditional on the reconstructed event,
from which the $\tau$ momenta and the polarimetric vectors follow.
Modeling the full conditional density rather than performing a point regression ensures that each
sampled neutrino configuration respects the correlations between the momentum components, and allows
the spin observables of each event to be estimated as averages over these samples, together with
their uncertainties.
The models take as inputs the four-momenta of the visible $\tau$ decay products, displacement
variables sensitive to the finite $\tau$ lifetime, and the missing transverse momentum, which are
combined by a transformer-based conditioning network.
The resolution achieved on the reconstructed spin observables significantly improves on that obtained from a
stand-alone transformer regressor trained with a mean-squared-error loss.

Using simulated LHC proton-proton collisions, including detector resolution effects,
we demonstrated the method in three applications.
Fitting the reconstructed spin correlation matrix and polarization vectors, we find that the presence of quantum entanglement in
$H\rightarrow\tau\tau$ decays can be distinguished from its absence with a significance of at least $4.3\sigma$, using
the integrated luminosity expected at the HL-LHC.
We find that the sensitivity to \CP violation in $H\rightarrow\tau\tau$ decays can be increased by at least 36\% using the \phicp reconstructed with the estimated polarimetric vectors.
\texttt{TauPolaris} additionally provides an event-level uncertainty on the reconstructed \phicp
from the neutrino posterior, offering the potential for a further improvement in sensitivity
through an optimized event selection, which we leave to future work.
Finally, we constructed variables from the reconstructed polarimetric vectors that separate
spin-0 Higgs boson events from spin-1 $Z$ boson events, and which are independent of the \CP state of the Higgs boson, providing an additional handle for suppressing $Z\rightarrow\tau\tau$ background in
Higgs boson searches.

\FloatBarrier
\begin{acknowledgments}
D.W. was supported by the Science and Technology Facilities Council [grant number ST/W000636/1].
L.R. was supported by a Schr\"{o}dinger Scholarship from Imperial College London.

\end{acknowledgments}

\smallskip

\section*{Author contributions}

D.W. conceived and led the project. L.R. proposed and implemented the transformers and the gradient-based MAP estimation. D.W. was responsible for the applications to quantum entanglement and background suppression. L.R. and D.W. contributed equally to all other elements of this work.

\bibliographystyle{apsrev4-1}
\bibliography{lit}
\end{document}